\documentclass[twocolumn]{aastex63}
\usepackage{amssymb, amsmath}
\usepackage{graphicx}
\usepackage{natbib}
\usepackage{latexsym}
\usepackage{mathrsfs}
\usepackage{mathtools}
\usepackage{multirow}
\usepackage{xcolor}
\usepackage{color, colortbl}
\usepackage{booktabs,makecell}
\usepackage{float}
\usepackage{textcase}

\newcommand{\tsup}{\textsuperscript}

\newcommand{\LHS}{\mbox{LHS 3844}}
\newcommand{\Lya}{Ly$\alpha$}

\begin{document}

\title{The high-energy spectrum of the nearby planet-hosting inactive mid-M dwarf \LHS}

\correspondingauthor{Hannah Diamond-Lowe}
\email{hdiamondlowe@space.dtu.dk}

\author[0000-0001-8274-6639]{Hannah Diamond-Lowe}
\affiliation{Center for Astrophysics $\vert$ Harvard \& Smithsonian, 60 Garden St., Cambridge, MA 02138, USA}
\affiliation{National Space Institute, Technical University of Denmark, Elektrovej, 2800 Kgs.\ Lyngby, Denmark}

\author[0000-0002-1176-3391]{Allison Youngblood}
\affiliation{Laboratory for Atmospheric and Space Physics, 1234 Innovation Dr., Boulder, CO 80303, USA}

\author[0000-0002-9003-484X]{David Charbonneau}
\affiliation{Center for Astrophysics $\vert$ Harvard \& Smithsonian, 60 Garden St., Cambridge, MA 02138, USA}

\author[0000-0002-3641-6636]{George King}
\affiliation{Department of Astronomy, University of Michigan, 1085 S.\ University Ave., 323 West Hall, Ann Arbor, MI 48109, USA}

\author[0000-0002-1912-3057]{D.\ J.\ Teal}
\affiliation{Department of Astronomy, University of Maryland, College Park, MD 20742, USA}

\author[0000-0003-2052-3442]{Sandra Bastelberger}
\affiliation{Department of Astronomy, University of Maryland, College Park, MD 20742, USA}

\author[0000-0002-5466-3817]{Lia Corrales}
\affiliation{Department of Astronomy, University of Michigan, 1085 S.\ University Ave., 323 West Hall, Ann Arbor, MI 48109, USA}

\author[0000-0002-1337-9051]{Eliza M.-R.\ Kempton}
\affiliation{Department of Astronomy, University of Maryland, College Park, MD 20742, USA}

\begin{abstract}

To fully characterize the atmospheres, or lack thereof, of terrestrial exoplanets, we must include the high-energy environments provided by their host stars. The nearby mid-M dwarf \LHS\ hosts a terrestrial world that lacks a substantial atmosphere. We present a time-series UV spectrum of \LHS\ from 1131 to 3215\AA\ captured by HST/COS. We detect one flare in the FUV that has an absolute energy of $8.96\pm0.77\times10^{28}$ erg and an equivalent duration of $355\pm31$ s. We extract the flare and quiescent UV spectra separately. For each spectrum, we estimate the \Lya\ flux using correlations between UV line strengths. We use Swift-XRT to place an upper limit on the soft X-ray flux and construct a differential emission model to estimate flux that is obscured by the interstellar medium. We compare the differential emission model flux estimates in the XUV to other methods that rely on scaling from the \Lya, Si \textsc{iv}, and N textsc{v} lines in the UV. The XUV, FUV, and NUV flux of \LHS\ relative to its bolometric luminosity is log$_{10}$($L_{\mathrm{band}}/L_{\mathrm{Bol}}$) = $-3.65$, $-4.16$, and $-4.56$, respectively, for the quiescent state. These values agree with trends in high-energy flux as a function of stellar effective temperature found by the MUSCLES survey for a sample of early-M dwarfs. Many of the most spectroscopically accessible terrestrial exoplanets orbit inactive mid-to-late M dwarfs like \LHS. Measurements of M dwarf high-energy spectra are preferable for exoplanet characterization but are not always possible. The spectrum of \LHS\ is a useful proxy for the current radiation environment for these worlds.\\
\end{abstract}


\section{Introduction} \label{sec:intro}

Studying the atmospheres of terrestrial exoplanets can provide insights into their histories and current processes that are not available from radius and mass measurements alone. The most accessible terrestrial planet atmospheres belong to worlds with high equilibrium temperatures ($T_{\mathrm{eq}}>400$ K) that transit small ($<0.3\ M_{\odot}$), nearby ($<15$ pc) stars. These systems provide large planet-to-star radius ratios $R_{\mathrm{p}}/R_{\mathrm{s}}$ and high signal-to-noise ratio observing opportunities. The Transiting Exoplanet Survey Satellite \citep[TESS; ][]{Ricker2015}, along with other ground-based survey programs \citep[e.g., ][]{Nutzman2008,Gillon2013,Irwin2015}, is expanding the known sample of transiting terrestrial exoplanets orbiting nearby mid-M dwarfs, and these worlds are becoming prime targets for in-depth characterization. 

Mid-M dwarfs have masses and radii that are typically 20\% times that of the Sun and exhibit stark differences from their G-type counterparts. M dwarfs have extended pre-main-sequence phases \citep{Baraffe2015} and remain chromospherically active on long timescales \citep{Newton2017}. A large fraction of M dwarf surfaces are covered by magnetic fields, which heat their upper atmospheres and produce strong emission lines and flaring in these fully convective stars \citep{Saar1985,Loyd2018,Wright2018}. For planets in orbit around young mid-M dwarfs, this adds up to a harsh high-energy environment that can alter their atmospheric chemistry and even strip their atmospheres altogether. Once a planet-hosting M dwarf settles onto the main sequence and slows its rotation (around 2 Gyr), it provides a more benign environment.

If a primordial planetary atmosphere survives the youthful stages of a mid-M dwarf host star, or the planet outgasses a secondary atmosphere later on, the high-energy M dwarf environment still plays a role in sculpting the planetary atmosphere. Molecular photodissociation cross sections are wavelength-dependent, so planetary atmospheric chemistry is highly sensitive to the stellar spectral energy distribution (SED). For instance, the far-ultraviolet (FUV = 912--1700 \AA) to near-ultraviolet (NUV = 1700--3200 \AA) flux ratio can shift the abundances of key molecules for habitability studies like water, methane, carbon dioxide, ozone, and diatomic oxygen by more than an order of magnitude \citep{Harman2015,Rugheimer2015,France2016}. While high-energy radiation may be damaging to biological material, there is evidence that some amount is needed for abiogenesis \citep{Ranjan2017}.

Studies of the atmospheric escape of terrestrial exoplanets orbiting M dwarfs are rooted in theoretical studies of what early Venus, Earth, and Mars might have looked like in the presence of a young, more active Sun \citep[e.g.,][]{Watson1981,Zahnle1990}. X-ray and extreme ultra-violet flux (X-ray = 1--100\AA; EUV = 100--912\AA; XUV = 1--912\AA) are responsible for hydrodynamic escape, by which photolysis of atmospheric molecules, namely water, can produce a stream of escaping hydrogen that can drag heavier molecules with it. It remains an open question whether or not terrestrial worlds around M dwarfs can retain atmospheres. Knowing where the divide is between those that can and those that cannot is a key piece of missing information when it comes to identifying optimal terrestrial targets for atmospheric follow-up. This divide, known as the cosmic shoreline, relies on many unknowns, including high-energy stellar flux levels over planetary lifetimes \citep{Zahnle2017}.

In this work, we focus on the nearby mid-M dwarf \LHS, which hosts a terrestrial exoplanet discovered by TESS \citep{Vanderspek2019}. \citet{Kreidberg2019} observed nine orbits of the terrestrial exoplanet \LHS b with 100 hr of Spitzer time. From the phase-curve data, they determined that surface pressures of 10 bars or higher are disfavored on this planet for a range of atmospheric compositions. Clear, low mean molecular weight atmospheres at surface pressures of 0.1 bars or greater are disfavored by optical ground-based transmission spectroscopy \citep{Diamond-Lowe2020b}. On the basis of stellar wind and high-energy flux over the planetary lifetime of \LHS b, \citet{Kreidberg2019} also disfavored tenuous high mean molecular weight atmospheres, with a small area of parameter space allowed if such an atmosphere can be continuously replenished. It appears that the terrestrial world \LHS b does not have an atmosphere, and in this paper, we quantify one likely contributor, namely the high-energy emission from its host star.

We do not know the high-energy history of \LHS, but we here provide a current snapshot as an anchor for stellar evolution and atmospheric escape models. In Section~\ref{sec:obs} we describe in detail our observations and resulting data. We present an analysis of the observed FUV flare in Section~\ref{sec:flare} and the steps to constructing a panchromatic spectrum for \LHS\ in Section~\ref{sec:spec}. We discuss the implications of our data for the terrestrial exoplanet \LHS b in Section~\ref{sec:disc}. We conclude with Section~\ref{sec:conclusion}. We provide a list of constants used throughout this work in Table~\ref{tab:constants}.

All of our data products are available as high-level science products (HLSPs) at
MAST via \dataset[10.17909/t9-fqky-7k61]{\doi{10.17909/t9-fqky-7k61}}.

\begin{deluxetable*}{l|ccc}
\centering
\caption{Constants for \LHS\ and \LHS b Used in This Work\label{tab:constants}}
\tablewidth{0pt}
\tablehead{
\colhead{Constant} & \colhead{Unit} & \colhead{Value} & \colhead{Reference}
}
\startdata
Distance & pc & $14.8909\pm0.0113$ & \citet{GaiaDR22018} \\
Stellar radius & $R_{\odot}$ & $0.178\pm0.012$ & \citet{Kreidberg2019}\\
Stellar mass & $M_{\odot}$ & $0.158\pm0.004$ & \citet{Kreidberg2019}\\
Effective temperature & K & $3036\pm77$ & \citet{Vanderspek2019}\\
Bolometric flux (at Earth) & erg cm$^{-2}$ s$^{-1}$ & $3.52\pm0.33\times10^{-10}$ & \citet{Kreidberg2019}\\
Spectral type & & M4.5--M5 & \citet{Vanderspek2019}\\
\hline
Planet radius & $R_\oplus$ & $1.244\pm0.006$ & \citet{Kreidberg2019}\\
Planet mass & $M_\oplus$ & $2.2\pm1.0$ & \citet{Chen2017}\tsup{a}\\
Semi-major axis & au & $0.00622\pm0.00017$ & \citet{Vanderspek2019}\\
Equilibrium temperature & K & $805\pm20$ & \citet{Vanderspek2019}\\
\enddata
\tablecomments{We primarily take values from the more recent \citet{Kreidberg2019} study. When values are taken from \citet{Vanderspek2019}, they are either fixed or not used in the \citet{Kreidberg2019} analysis.\\
\tsup{a}We use the \citet{Chen2017} relation to estimate the mass of \LHS b.}
\end{deluxetable*}

\section{Observations} \label{sec:obs}

We observed the UV spectrum of the mid-M dwarf \LHS\ ($V=15.3$, $K=9.1$) from 1131 to 3215 \AA\ with 10 orbits of the Hubble Space Telescope (HST) and the Cosmic Origins Spectrograph (COS; GO Program 15704; PI: Diamond-Lowe). The 10 orbits were divided into three visits executed from 2019 August 7 to 2019 August 12. In order to cover the full wavelength range, we use three COS gratings: G130M (centered at 1291 \AA), G160M (1600 \AA), and G230L (2950 \AA). It is not possible to observe with multiple COS gratings simultaneously. We spent six orbits observing with the G130M grating, three with G160M, and one with G230L. 

\begin{figure*}
\includegraphics[width=\textwidth]{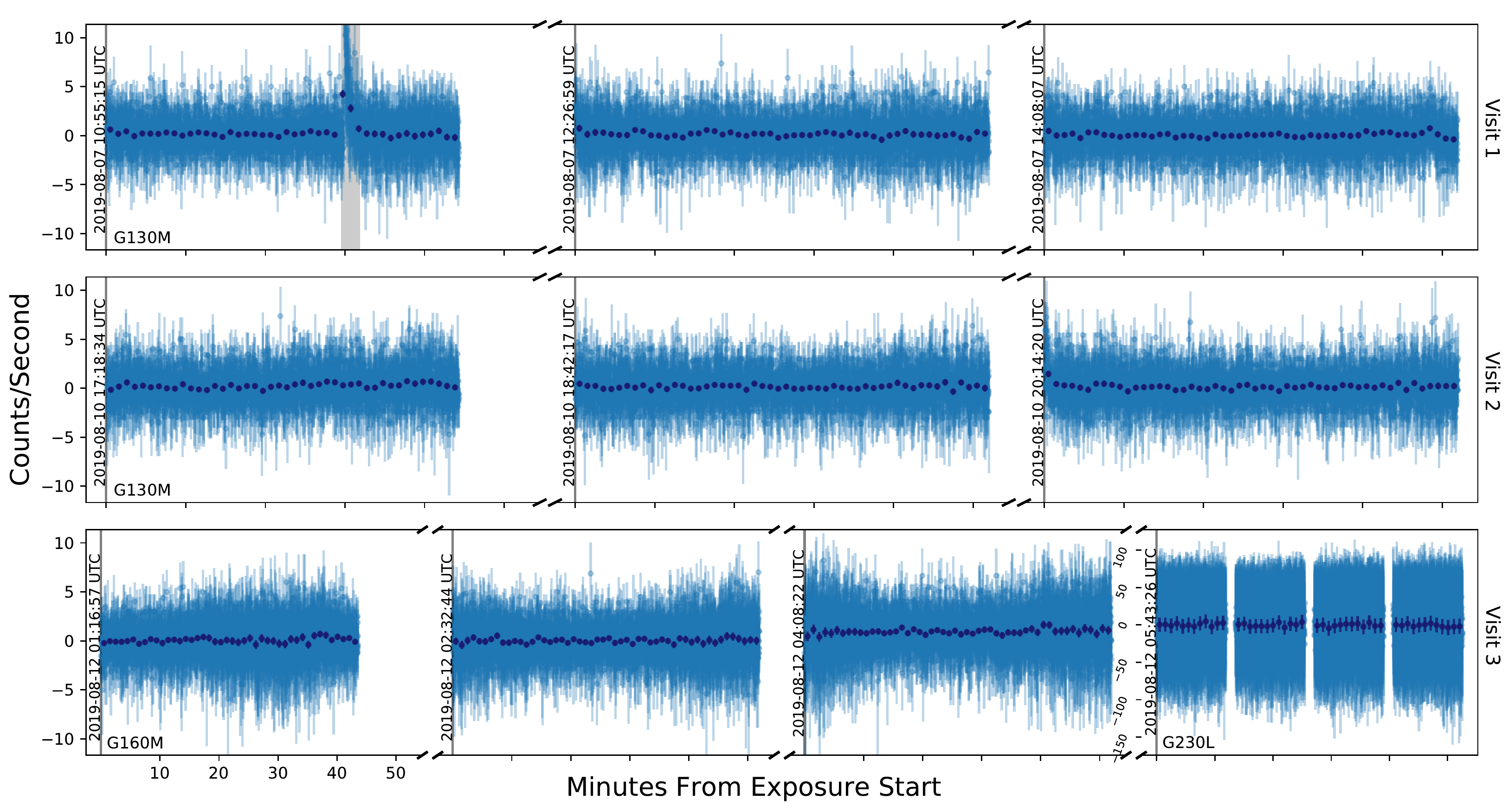}
\caption{Time series of observations of \LHS\ taken with HST/COS for GO Program 15704 (PI: Diamond-Lowe). Ten orbits of HST time were divided into three visits. The start times of each exposure are indicated with a vertical line and labeled in UTC. Blue data points with 1$\sigma$ Poisson error bars are counts per second in 1 s time bins. Dark blue points are the same but binned into 1 minute time bins. The $x$-axis measures the minutes from the start of the exposure; all $x$-axis tick marks are at equal intervals of 10 minutes and labeled only once in the bottom left corner for clarity. Note the change in the $y$-axis scale for the G230L exposure. The first exposure of every visit is shorter to allow time for target acquisition before the start of the exposure. A gray band in the first G130M exposure indicates the flare.}
\label{fig:timeseries}
\end{figure*}

COS has two separate detectors for observing the FUV (G130M and G160M) and NUV (G230L), while the gratings all have different throughputs. The dark current in the NUV detector is 2 orders of magnitude greater than that of the FUV detector, leading to noisier NUV count rates. There is, however, detectable stellar continuum flux in the NUV that is not present in the FUV. There is a limit to the signal-to-noise ratio achievable with each grating due to fixed-pattern noise in the COS detectors. This can be overcome by changing the fixed-pattern position (\texttt{FP-POS}) between exposures. Each \texttt{FP-POS} shifts the spectrum slightly in the dispersion direction so that the spectral features fall on different parts of the detector. The \texttt{FP-POS} is changed during the Earth occultations for the G130M and G160M gratings. For the G230L grating, it is changed during the final orbit, hence the split into four separate exposures with short data gaps (Figure~\ref{fig:timeseries}). The allocation of the orbits by grating was designed to achieve a signal-to-noise ratio of at least 4 in prominent UV lines. A UV spectrum of the mid-M dwarf star GJ 1214 was scaled to the distance and size of \LHS\ in order to estimate the exposure times needed in each grating \citep{France2016}. Table~\ref{tab:obs} provides details about each observation.

We used the G130M and G160M gratings to observe prominent molecular lines in the FUV and the G230L grating to observe lines in the NUV. We used the \texttt{FLASH=YES} setting and \texttt{TIME-TAG} mode in order to maximize the exposure time on \LHS\ and create a time series in which to look for stellar variability and flares. This observing strategy was based on the successful Measurements of the Ultraviolet Spectral Characteristics of Low-mass Exoplanetary Systems (MUSCLES; GO Program 13650) and Mega-MUSCLES (GO Program 15071) treasury surveys \citep{France2016,Froning2019}. 

\begin{deluxetable*}{cccccccc}
\centering
\caption{Observations of \LHS\ with HST/COS\label{tab:obs}}
\tablewidth{0pt}
\tablehead{
\colhead{Visit} & \colhead{Orbit} & \colhead{Grating} & \colhead{Central Wavelength} & \colhead{Wavelength Range} & \colhead{Resolution Range} & \colhead{Exposure Time\tsup{a}} & \colhead{\texttt{FP-POS}} \\
\colhead{} & \colhead{}      & \colhead{}             & \colhead{(\AA)}        & \colhead{(\AA)} & \colhead{($\times10^3$)} & \colhead{(s)} &
}
\startdata
  & 1  &       &      &            &        & 2661 & 3 \\
1 & 2  & G130M & 1291 & 1131--1429 & 12--16 & 3120 & 3\\
  & 3  &       &      &            &        & 3120 & 4 \\
\hline
  & 4  &       &      &            &        & 2661 & 3\\
2 & 5  & G130M & 1291 & 1131--1429 & 12--16 & 3120 & 4 \\
  & 6  &       &      &            &        & 3120 & 4\\
\hline
\multirow{4}{*}{3} & 7  &       &      &            &        & 2614 & 2 \\
 & 8  & G160M & 1600 & 1407--1775 & 13--20 & 3120 & 3\\
 & 9  &       &      &             &        & 3120 & 4 \\
 & 10 & G230L & 2950 & 1678--3215 & 2.1--3.9 & 4$\times$ 716  & 1, 2, 3, 4  \\
\enddata
\tablecomments{Spectra taken with the G130M and G160M gratings were placed at lifetime position four (LP4) on the FUV detector in accordance with recommendations by the COS team to mitigate gain sag and extend the detector lifetime.\\
\tsup{a} The first exposure of every visit is shorter to allow for time to acquire the target.\\
 }
\end{deluxetable*}

\section{Time Series of \LHS}\label{sec:flare}

The star \LHS\ is considered inactive, with an H$\alpha$ equivalent width of $0.12\pm0.08$ \AA\ and one detected optical flare at an energy of $1.30\times10^{31}$ erg in the TESS bandpass, observed at a 2-minute cadence by TESS in a 27 day sector \citep{Medina2020}. However, M dwarfs are known to flare at high energies even when relatively quiescent at optical wavelengths \citep{Loyd2018}. We use the \texttt{corrtag} files associated with each HST/COS exposure to construct a time series of \LHS\ (Figure~\ref{fig:timeseries}). We detect one flare in our time series, observed with the G130M grating. We note that we could have detected a flare in any of the orbits (in any of the gratings) but spent the most time observing with G130M. We also note that the additional noise in the G160M and G230L gratings means that the energy threshold for detecting flares is higher in these gratings than in the G130M grating.

\subsection{Flux-calibrating the time series}

The \texttt{corrtag} files provide a count rate (counts per second) that we need to flux-calibrate in order to calculate the flare properties, such as FWHM, peak flux, time of peak flux, absolute energy, and equivalent duration. This is not directly provided in the COS data products, so we start with the provided \texttt{corrtag} files and manually perform the background subtraction detailed in chapter 3.4.19 of the COS Data Handbook, version 4.0\footnote{\href{https://hst-docs.stsci.edu/cosdhb/chapter-3-cos-calibration/3-4-descriptions-of-spectroscopic-calibration-steps\#id-3.4DescriptionsofSpectroscopicCalibrationSteps-3.4.19BACKCORR:1DSpectralBackgroundSubtraction}{COS Data Handbook v4.0, Chapter 3.4.19}}. Then, following the work of \citet{Loyd2014} and \citet{Loyd2018}, we use the \texttt{x1d} (spectral) data products, which include information about the wavelength grid, calibrated flux, and net counts to back out an effective area ($A_\mathrm{eff}$) for the COS detector. This allows us to include the COS detector response in the time-series flux calibration.

Every detected photon event from the \texttt{corrtag} files has an associated wavelength. We use this wavelength to calculate the photon energy and then divide by the associated effective area. Finally, we bin these events in time in order to achieve a time series of fluxes in erg cm$^{-2}$ s$^{-1}$. We assume Poisson uncertainties for each counted photon and propagate these values though our flux-calibration process to get the uncertainties on the fluxes in the time series.

\subsection{Absolute Energy and Equivalent Duration}

Following \citet{Davenport2014}, \citet{Hawley2014}, and \citet{Loyd2018}, we compute the absolute FUV flare energy $E$ as

\begin{equation}
    E = 4\pi d^2 \int_{\mathrm{flare}} (F_f - F_q) dt
\end{equation}

\noindent where $d$ is the distance to the star, $F_f$ is the flux during the flare, and $F_q$ is the estimated quiescent flux. A related quantity, the equivalent duration $\delta$, is calculated as

\begin{equation}
    \delta = \int_{\mathrm{flare}} \frac{F_f - F_q}{F_q} dt
\end{equation}

\noindent which is essentially the area under the flare.

In order to calculate the absolute energy and equivalent duration of the flare, we need to (1) choose a flare duration to integrate over and (2) estimate the quiescent flux level. We determine which points are included in the flare by shifting the starting and ending points of the flare and then calculating the flare energy at each shift. When shifting the start time of the flare, we keep the end time constant, and vice versa. Once the energy is stable to the change in flare duration over which we integrate, we have found the extent in time of the flare. The points that we include in the flare range from 1783 to 1916 s after the start of the exposure, giving a flare duration of 133 s. To estimate the quiescent flux, we fit a quadratic function in time to the out-of-flare flux. We present the flux-calibrated time series that includes the observed flare in Figure~\ref{fig:flare}, with the in-flare points highlighted in orange and the estimate of the quiescent flux shown as a gray line.

Dedicated analyses of flare surveys perform flare injection-and-recovery tests in order to estimate the completeness of flare detections for different energy thresholds, as well as place uncertainties on the flare properties \citep{Loyd2018,Medina2020}. Flare injection-and-recovery tests are outside the scope of this work. We instead estimate the uncertainties in absolute energy and equivalent duration of our single flare using a Monte Carlo estimate. We shuffle each in-flare flux point around by an amount drawn from a normal distribution centered on zero with a standard deviation equal to the uncertainty in the flux. We do this 1000 times and at each iteration calculate the absolute energy and equivalent duration. We take the standard deviations of the resulting distributions of these values as their uncertainties (Table~\ref{tab:flareparams}). Needless to say, this is not a Bayesian approach, and our uncertainties mostly capture decisions regarding the quiescent flux level and flare duration.

\subsection{FWHM, Peak Time, and Amplitude}

We use the classical flare model from \citet{Davenport2014} to fit for the FWHM,\footnote{Here FWHM is defined as the time span, including the rise and decay portions of the flare, at which the flux is half of the maximum recorded flux \citep{Kowalski2013}.} the time of peak flux, and the amplitude of the flare. From this classical flare model, we can also do a by-eye check for the starting and ending times of the flare that we chose in the previous section. The \citet{Davenport2014} flare model assumes that the quiescent state is at zero, so we subtract off the quadratic fit to the quiescent flux before fitting the flare model.

\begin{figure}
\includegraphics[width=0.4725\textwidth]{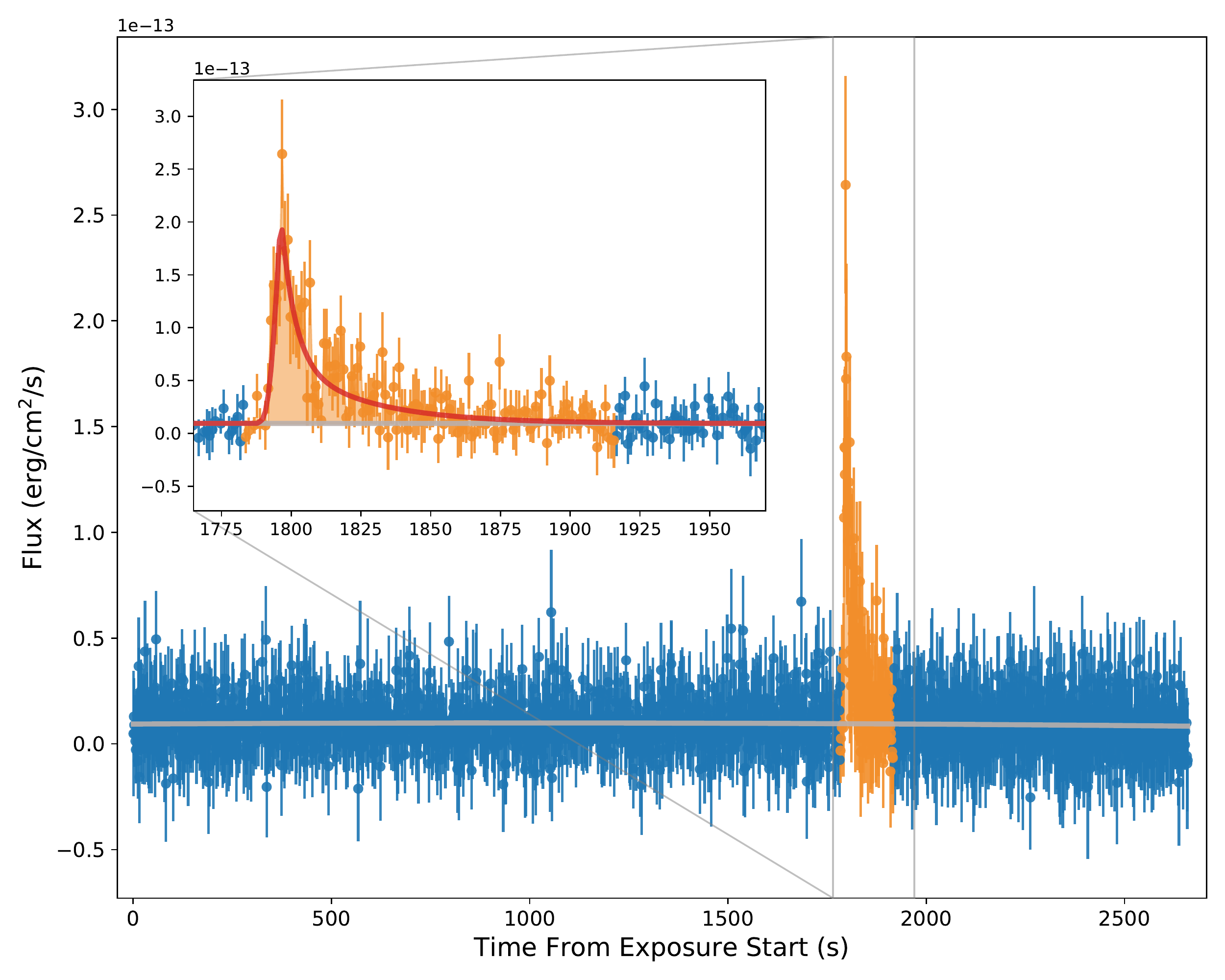}
\caption{Flux-calibrated time series of the first G130M exposure of visit 1 (GO Program 15704). Blue points with 1$\sigma$ error bars are flux-calibrated using the COS \texttt{corrtag} and \texttt{x1d} files. Orange points highlight the flare, from 1783 to 1916 s after the start of the exposure. Time bins are 1 s. The gray line is a quadratic fit to the out-of-flare data, which we take as the quiescent flux level. The red line in the inset is a fit of the classic flare model to the data \citep{Davenport2014}. The shaded orange region between the flare and quiescent flux is the equivalent duration $\delta$.}
\label{fig:flare}
\end{figure}

We use the open-source fitting package \texttt{lmfit} \citep{Newville2016}, which includes a wrapper for the Markov Chain Monte Carlo (MCMC) package \texttt{emcee} \citep{Foreman-Mackey2013}. When fitting the flare model to the data, we weight the data by the flux uncertainties. We use 100 walkers and run the MCMC for 5000 steps with a 300 step burn-in, at which point the chain length is over 50 times the integrated autocorrelation time. The final fit for the \citet{Davenport2014} flare model is presented as a red line in the inset of Figure~\ref{fig:flare}, with the associated values listed in Table~\ref{tab:flareparams}. 

\begin{deluxetable}{l|cc}
\centering
\caption{Properties of the FUV Flare from \LHS\label{tab:flareparams}}
\tablewidth{0pt}
\tablehead{
\colhead{} & \colhead{Unit} & \colhead{Value}
}
\startdata
Wavelength range & \AA & 1131--1429 \\
Flare start and end    & s & 1783--1916\\
Flare duration         & s & 133\\
Mean quiescent flux & erg cm$^{-2}$ s$^{-1}$ & $9.48\times10^{-15}$\\
\hline
Absolute energy ($E$)  & erg & 8.96 $\pm$ 0.77 $\times$ 10$^{28}$ \\
Equivalent duration ($\delta$) & s   & 355 $\pm$ 31\\
\hline
FWHM                & s   & 7.9 $\pm$ 1.0\\
Time of peak flux   & s   & 1796.27 $\pm$ 0.48\\
Amplitude           & erg cm$^{-2}$ s$^{-1}$ & 1.99 $\pm$ 0.24 $\times$ 10$^{-13}$
\enddata
\tablecomments{This flare was detected with the G130M grism, which captured flux in the range of 1131--1429\AA. Values reported here represent that whole flux range.}
\end{deluxetable}

\subsection{Flare Rate}

From a total of 17,802 s of observing time with the G130M grating, we can place a rough upper limit on the FUV flare frequency of \LHS\ at $4.9\pm2.2$ flares day$^{-1}$, assuming Poisson counting statistics and not accounting for the noncontiguous nature of the G130M observations. The absolute energy and equivalent duration are helpful in contextualizing the \LHS\ flare against other flaring M dwarfs. We find that if we take the upper limit flare frequency from our detection, the values we calculate for $E$ and $\delta$ are within the distribution uncertainties of inactive M stars in the MUSCLES sample \citep[Figure 3 of ][]{Loyd2018}.

Using TESS data, \citet{Medina2020} detected one \LHS\ flare and reported a flare frequency of $2.4\pm2.0\times10^{-4}$ flares day$^{-1}$ at or above energies of $3.16\times10^{31}$ erg in the TESS bandpass for \LHS. TESS operates in the optical, so we cannot draw direct comparisons between the flare frequency reported by \citet{Medina2020} and the upper limit we place here because we do not know the bolometric flux of these flare events. There are no concurrent observations between TESS and HST/COS to determine if the FUV flare we detect had an observable optical counterpart. However, \citet{Medina2020} found that the mid-M dwarfs in their sample follow a common flare energy distribution with a power-law index of $-2$; i.e., low-energy flares are more common than high-energy flares. If we can assume that the bolometric flux of the low-energy flare we report here is lower than the bolometric flux of the flare detected by \citet{Medina2020} in the TESS data, it is possible that low-energy flares like the one we observe are actually fairly common on \LHS. This would also be in line with the finding that even stars that are considered inactive (EW$_{\mathrm{Ca\ \textsc{ii}}}<2$\AA) exhibit flares in both the optical \citep{Medina2020} and the UV \citep{Loyd2018}. 

\section{Panchromatic Spectrum of \LHS} \label{sec:spec}

Given the presence of a flare in the UV data, we construct two panchromatic spectra for \LHS: a flare spectrum and a quiescent spectrum. However, data that include the flare were only observed in the FUV with the G130M grating, so the flare spectrum is a much rougher estimate of what the \LHS\ flux might look like during a relatively low-energy flare. The panchromatic spectra we present from 1 to $1\times10^5$\AA\ is constructed from (1) Swift-XRT soft X-ray data, (2) estimates of the EUV and \Lya\ flux, (3) HST/COS data in the FUV and NUV, (4) a PHOENIX model in the optical to mid-infrared, and (5) a blackbody tail out to 10 $\mu$m.

\subsection{Separating the HST/COS quiescent and flare spectra}

In order to excise the flare and its associated spectra from the G130M time series, we use the \texttt{costools}\footnote{\href{https://github.com/spacetelescope/costools}{github.com/spacetelescope/costools}} \texttt{splittag} function to create three new sets of \texttt{corrtag} files for the first G130M exposure: one before the flare, one during the flare, and one after the flare. We then use the \texttt{x1dcorr} function to get flux-calibrated spectra from each \texttt{corrtag} set. These flux-calibrated spectra are akin to the \texttt{x1d} data products provided by the HST/COS pipeline. To turn our homemade \texttt{x1d} files into \texttt{x1dsum} files that combine multiple exposures taken with the same grating, we follow the instructions from chapter 3.4.22 of the COS Data Handbook version 4.0,\footnote{\href{https://hst-docs.stsci.edu/cosdhb/chapter-3-cos-calibration/3-4-descriptions-of-spectroscopic-calibration-steps\#id-3.4DescriptionsofSpectroscopicCalibrationSteps-3.4.22Finalization(makingthex1dsumfiles)}{COS Data Handbook v4.0, Chapter 3.4.22}} which explains how to sum \texttt{x1d} files by weighting by the data quality (\texttt{DQ\_WGT}). In this way, we combine all nonflare (quiescent) portions of the G130M exposures and leave the flare on its own. 

The G160M and G230L exposures are only applicable to the quiescent spectrum. The G230L grating does not provide complete coverage of the NUV; there is a gap in the data between 2125 and 2750\AA. We take the average flux out to 300\AA\ on either side of the gap and take this as the nominal flux value in the gap. There are no emission lines that fall in the data gap that would have been visible over the data noise. A list of the emission lines that we measure, along with which gratings we use and the total exposure times to measure them, can be found in Table~\ref{tab:measuredemission}.

\begin{deluxetable*}{c|ccccccc}
\centering
\caption{Measured Emission Lines from \LHS\ with HST/COS\label{tab:measuredemission}}
\tablewidth{0pt}
\tablehead{
\colhead{} & \colhead{Grating} & \colhead{Total Exposure Time} & \colhead{Line} & \colhead{Line Centers} & \colhead{log$_{10}$(Surface Flux)\tsup{a}} & \colhead{log$_{10}$(Surface Flux)\tsup{b}} & log($T$)\tsup{d}\\
\colhead{} & \colhead{}      & \colhead{(s)}                 &                & \colhead{(\AA)}        & \multicolumn{2}{c}{(erg cm$^{-2}$ s$^{-1}$)} & 
}
\startdata
\multirow{9}{*}{\rotatebox[origin=c]{90}{FUV}} & \multirow{7}{*}{G130M} & \multirow{7}{*}{17,802} & C \textsc{iii} & 1175.59\tsup{c} & 3.22 $\pm$ 0.22 & 5.05 $\pm$ 0.20 &4.8\\
                    &       &       & Si \textsc{iii} & 1206.50 & 2.95 $\pm$ 0.06 & 4.63 $\pm$ 0.16 & 4.7\\
                    &       &       & O \textsc{v}    & 1218.34 & 3.45 $\pm$ 0.03  & ------ & 5.3\\
                    &       &       & N \textsc{v}     & 1238.82, 1242.8060  & 3.30 $\pm$ 0.05 & 4.49 $\pm$ 0.18 & 5.2\\
                    &       &       & Si \textsc{ii}   & 1264.74 & 2.13 $\pm$ 0.14 & ------ &4.5\\
                    &       &       & C \textsc{ii}    & 1334.53, 1335.67 & 3.32 $\pm$ 0.22 & 4.59 $\pm$ 0.18 & 4.5\\
                    &       &       & Si \textsc{iv}   & 1393.72, 1402.74 & 3.07 $\pm$ 0.21 & 4.81 $\pm$ 0.14 &4.9\\
                    \cline{2-8}
                    & \multirow{2}{*}{G160M} & \multirow{2}{*}{8854} & C \textsc{iv} & 1548.19, 1550.78 & 3.97 $\pm$ 0.05 & ------ & 5.0\\
                    &       &       & He \textsc{ii}   & 1640.4 & 3.06 $\pm$ 0.54 & ------ & 4.9\\
\cline{1-8}
\multirow{2}{*}{\rotatebox[origin=c]{90}{NUV}} & \multirow{2}{*}{G230L} & \multirow{2}{*}{2864}   & \multirow{2}{*}{Mg \textsc{ii}}   & \multirow{2}{*}{2796.35} & \multirow{2}{*}{4.19 $\pm$ 0.98} & \multirow{2}{*}{------} & \multirow{2}{*}{4.5}\\
 &&&&\\
 \enddata
 \tablecomments{Surface fluxes (erg cm$^{-2}$ s$^{-1}$) are calculated by scaling the observed flux by the distance $d$ and stellar radius $R_{\mathrm{s}}$ of \LHS: $F_{\mathrm{Surf}} = F_{\mathrm{Obs}} \times\ (d/R_{\mathrm{s}})^2 $. We use $d = 14.8909 \pm 0.0113$ pc and $R_{\mathrm{s}} = 0.178 \pm 0.012\ R_{\odot}$ \citep{GaiaDR22018,Kreidberg2019}. For double lines, we present the combined surface flux. Some lines in the G130M grating are not well constrained by the data during the flare due to a low signal-to-noise ratio so we do not report them here. We do not have G160M or G230L observations during the flare. \\
 \tsup{a} Quiescent spectrum \\
 \tsup{b} Flare spectrum\\
 \tsup{c} There are six unresolved C \textsc{iii} transition lines here; we fit them as a single broad line\\
 \tsup{d} Peak formation temperatures from CHIANTI v7.0 \citep{Landi2012}
 }
\end{deluxetable*}

\subsection{\Lya\ estimation}

The largest flux source in the UV is \Lya, so having a measurement or estimation of the \Lya\ flux is necessary to construct the UV spectrum of \LHS. However, for all stars other than the Sun, \Lya\ emission is absorbed by neutral hydrogen in the interstellar medium (ISM) before it can reach our telescopes. The MUSCLES survey was able to gather enough flux in the wings of the \Lya\ profiles of early-M stars to perform a reconstruction of the line and thereby measure the flux \citep{Youngblood2016}. We make a distinction between the MUSCLES method of \Lya\ reconstruction, and the method we present here for \Lya\ estimation.

An additional barrier to measuring the \Lya\ flux from \LHS\ comes from the COS instrument. COS is optimal for measuring FUV flux \citep{Green2012,France2013}, but because it is a slitless spectrograph, the \Lya\ line becomes contaminated by geocoronal emission, whereby solar photons interact with neutral hydrogen in Earth's atmosphere and produce local \Lya\ emission. The Space Telescope Imaging Spectrograph (STIS) on board HST has a slit and can measure the wings of the \Lya\ profile \citep{France2016,Youngblood2016}, but at a lower efficiency. Given how faint \LHS\ is in the UV, we found that it would take a prohibitive amount of HST/STIS time (70 orbits) to build up enough signal-to-noise ratio in the wings of the \Lya\ line to perform a reconstruction. We remove contamination from the geocoronal \Lya\ emission, as well as the local emission of N \textsc{i} (1200\AA) and O \textsc{i} (1302, 1305, 1306\AA), from the HST/COS UV spectra by setting the flux values at the contaminated wavelengths to zero.

To estimate the \Lya\ flux from \LHS, we build upon the work of the MUSCLES survey to establish correlations between UV fluxes of different emission lines across their stellar sample. Though the early-M stars in the MUSCLES sample have orders-of-magnitude variations in their measured line fluxes, there exists statistically significant correlations between line fluxes within a given star's UV spectrum \citep{Youngblood2017}. Because the MUSCLES survey successfully reconstructed the \Lya\ flux for their sample and measured the UV--UV correlations between line strengths, including the \Lya\ line, we can build upon this work in order to estimate the \Lya\ flux from other measured UV lines.

\begin{figure*}
\includegraphics[width=\textwidth]{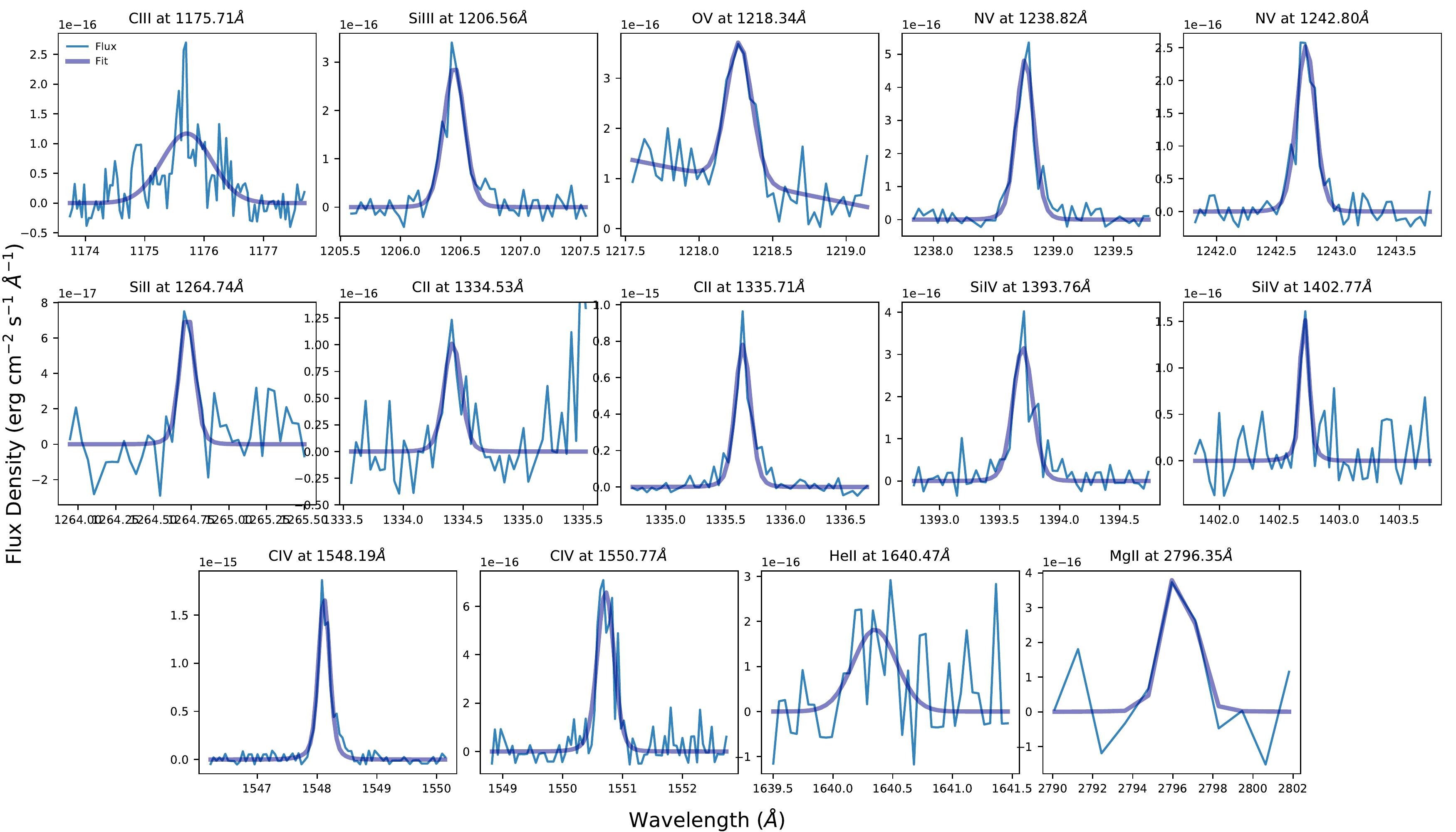}
\caption{Quiescent spectrum emission lines (thin light blue lines) listed in Table~\ref{tab:measuredemission}, with subplots for each transition. The transition lines are fitted with Gaussian functions convolved with COS LSFs (thick dark blue lines).}
\label{fig:lineprofiles_quiescent}
\end{figure*}

\begin{figure*}
\includegraphics[width=\textwidth]{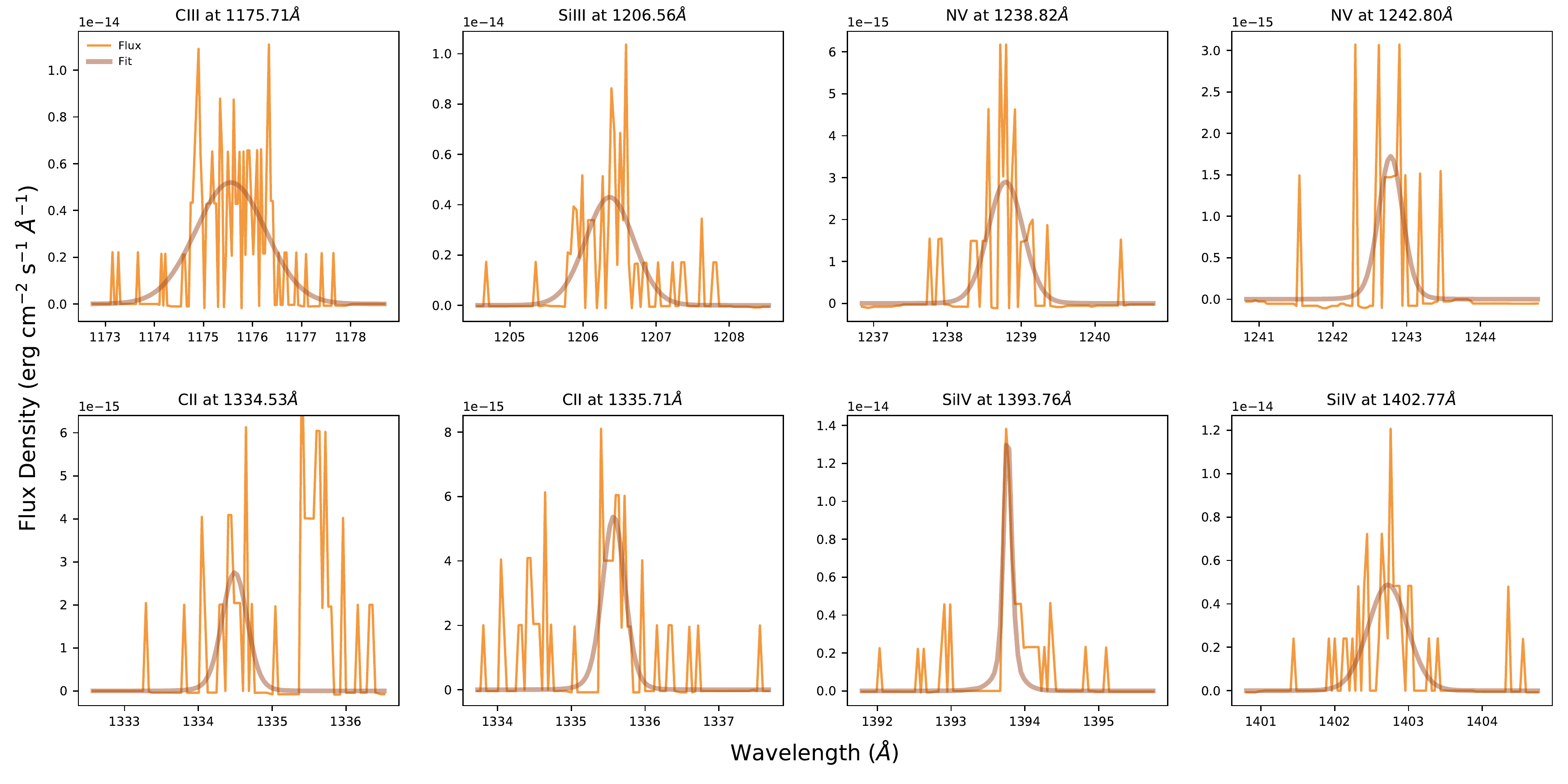}
\caption{Same as Figure~\ref{fig:lineprofiles_quiescent} but for the flare spectrum. Thin orange lines are the observed flux density during the flare. Thick brown lines are the fitted Gaussian functions convolved with COS LSFs. Note the difference in the \textit{x}- and \textit{y}-axes, as well as the lower signal-to-noise ratio.}
\label{fig:lineprofiles_flare}
\end{figure*}

We start by fitting Gaussian functions\footnote{We tried to fit Voigt functions to the UV lines, which is theoretically the correct function to use, but because our data have low signal-to-noise ratios, the extra free parameter made for an indistinguishable fit but with larger uncertainties.} convolved with the appropriate COS line spread function (LSF)\footnote{\href{https://www.stsci.edu/hst/instrumentation/cos/performance/spectral-resolution}{COS LSF}} to each of the emission lines listed in Table~\ref{tab:measuredemission}. Visuals of these fits are shown in Figures~\ref{fig:lineprofiles_quiescent} and~\ref{fig:lineprofiles_flare}, where we also separate out each transition. During the flare, the integrated flux in measured G130M emission lines increased by a median factor of 55, with no detectable trend with peak formation temperature that would point to which region of the stellar atmosphere the flare originated from. Indeed, other work finds that it is additional continuum flux that accounts for the overall rise in stellar radiation during flares \citep{Osten2015}. There was too little signal-to-noise ratio in the short time span of the flare to measure O \textsc{v} and Si \textsc{ii} emission.

\begin{figure}
\includegraphics[width=0.48\textwidth]{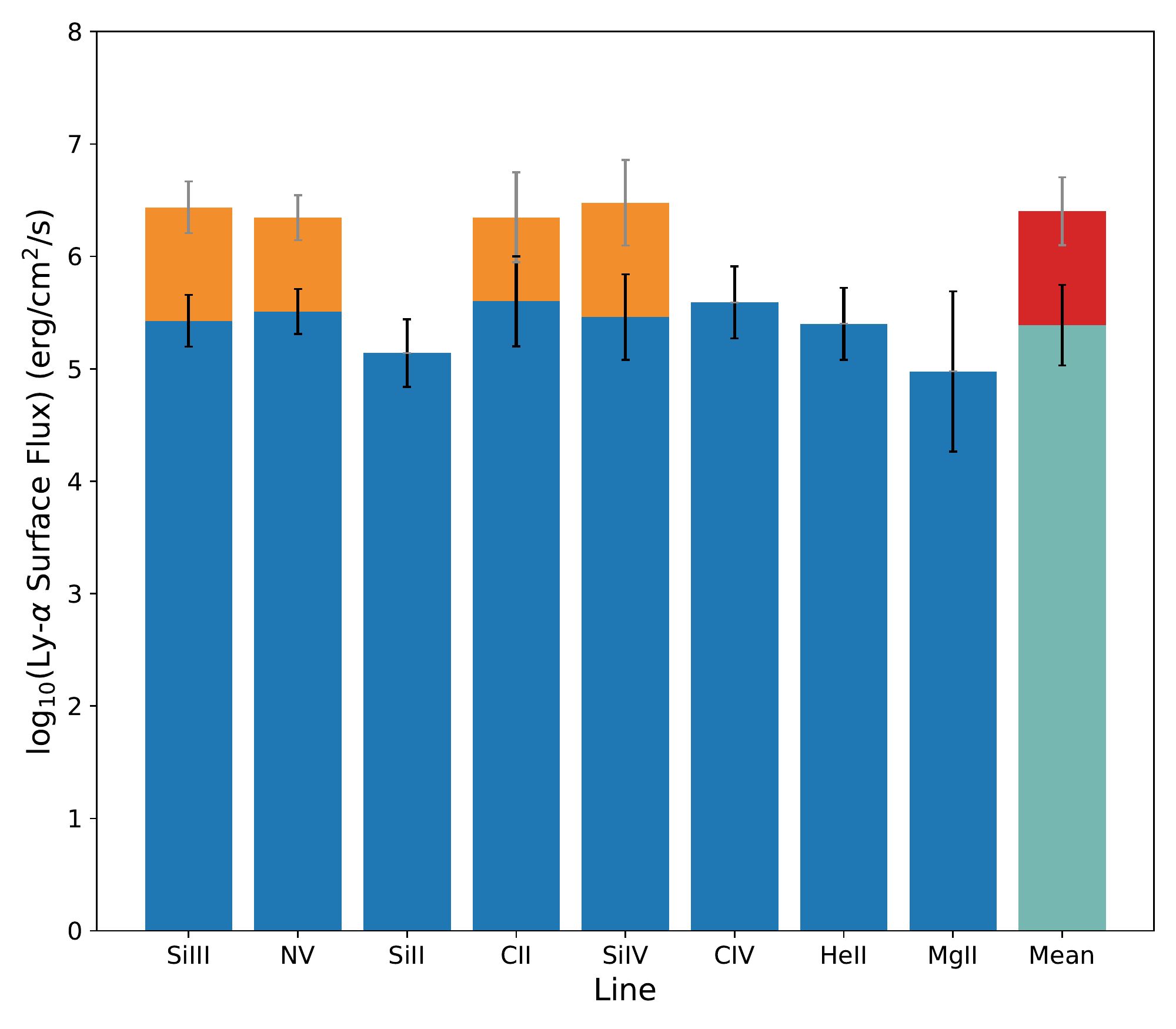}
\caption{The \Lya\ flux estimates from UV--UV line emission correlations found by \citet{Youngblood2017}. Blue and teal bars with black error bars are derived from the quiescent spectrum. The increase in surface flux during the flare is shown cumulatively with the orange and red bars and gray error bars. We take the mean of the individual line correlations and their respective standard deviations to get our final \Lya\ flux estimates.}
\label{fig:Lyaestimates}
\end{figure}

We integrate under the fitted Gaussians to establish the line flux. We then normalize the observed flux in each line to the surface flux of the star using the stellar distance and radius and take the base-10 log. In this format, we can use the UV--UV emission line relations established by \citet{Youngblood2017} to estimate this value for \Lya\ according to
\begin{equation}\label{eqn:UVscaling}
   \mathrm{log}_{10}(F_{\mathrm{Surf},\mathrm{ UV_1}}) = m \times \mathrm{log}_{10}(F_{\mathrm{Surf},\mathrm{ UV_2}}) + b
\end{equation}
\noindent where $F_{\mathrm{Surf},\mathrm{ UV_1}}$ in this case is the \Lya\ surface flux, $F_{\mathrm{Surf},\mathrm{ UV_2}}$ is the surface flux of the lines we measure (Table~\ref{tab:measuredemission}), and $m$ and $b$ are taken from Table 9 of \citet{Youngblood2017}. Note that not all emission lines we present in the \LHS\ UV spectrum are included in the UV--UV scalings, so these are not included in the \Lya\ estimate. To estimate the uncertainty in log$_{\mathrm{10}}$ \Lya\ surface flux, we compare the rms scatter about the best-fit lines of the UV--UV scalings \citep[][Table 9]{Youngblood2017} to the upper and lower values we get by calculating Equation~\ref{eqn:UVscaling} with 1$\sigma$ upper and lower bounds of the log$_{\mathrm{10}}$ surface flux of the measured UV lines. We take whichever value is higher as the uncertainty in the estimated log$_{\mathrm{10}}$ \Lya\ surface flux. In all cases, the rms values from \citet{Youngblood2017} are higher, meaning that the uncertainties in our \Lya\ estimates are dominated by the scatter in the UV--UV scalings, not by our measurements of emission lines in the COS data. We take the mean of the individual line estimates and their standard deviations to get our final \Lya\ flux estimates and uncertainties (Figure~\ref{fig:Lyaestimates}) and present the values in Table~\ref{tab:Lyaestimates}. 

\begin{deluxetable}{c|cc}
\centering
\caption{\Lya\ Surface Flux Estimates from Measured UV Lines \label{tab:Lyaestimates}}
\tablewidth{0pt}
\tablehead{
\colhead{Emission} &  \colhead{log$_{10}$($F_{\mathrm{Surf}, \mathrm{Ly}\alpha}$)\tsup{a}} & \colhead{log$_{10}$($F_{\mathrm{Surf}, \mathrm{Ly}\alpha}$)\tsup{b}} \\
\colhead{Line} & \colhead{erg cm$^{-2}$ s$^{-1}$} & \colhead{erg cm$^{-2}$ s$^{-1}$}
}
\startdata
Si \textsc{iii} & $5.43\pm0.23$ & $6.44\pm0.23$ \\
N \textsc{v}    & $5.51\pm0.20$ & $6.34\pm0.20$\\
Si \textsc{ii}  & $5.14\pm0.30$ & --- \\
C \textsc{ii}   & $5.60\pm0.40$ & $6.35\pm0.40$ \\
Si \textsc{iv}  & $5.46\pm0.38$ & $6.48\pm0.38$ \\
C \textsc{iv}   & $5.59\pm0.32$ & --- \\
He \textsc{ii}  & $5.40\pm0.32$ & --- \\
Mg \textsc{ii}  & $4.98\pm0.71$ & --- \\
\hline
Mean  & $5.39\pm0.36$ & $6.40\pm0.30$\\
\enddata
\tablecomments{Some lines that we measure, like C \textsc{iii} and O \textsc{v}, are not included in the UV--UV scaling relations from \citet{Youngblood2017}. \\
\tsup{a} Quiescent spectrum.\\
\tsup{b} Flare spectrum; only lines measured during the flare are used to estimate the \Lya\ surface flux (erg cm$^{-2}$ s$^{-1}$) from the flare spectrum.}
\end{deluxetable}

\subsection{EUV Estimation}\label{subsec:EUVestimation}

For stars other than the Sun, EUV (100--912 \AA) is not observable with any current telescope. Energies corresponding to 912\AA\ and below can ionize neutral hydrogen in the ISM, with energies approaching 912\AA\ the most likely to be absorbed. This makes detecting photons at 912\AA\ nearly impossible for most stars. The only dedicated EUV mission, the Extreme-Ultraviolet Explorer \citep[EUVE;][]{Craig1997}, which flew from 1992 to 2001, could measure emission under 600\AA\ for most nearby stars. EUVE captured spectra of several active M dwarfs, including AU Mic, AD Leo, EV Lac, and Proxima Centauri. Another high-energy observatory, the Far Ultraviolet Spectroscopic Explorer \citep[FUSE;][]{Moos2000,Sahnow2000}, operated from 1999 to 2007 and captured FUV spectra from 900 to 1200\AA. EUVE's all-sky survey did not pick up any flux at \LHS's coordinates, and FUSE did not make any observations within 1$^\circ$ of \LHS.

We present two methods for estimating the EUV flux of \LHS. The first relies on scaling relations similar to those used in the previous section to estimate the \Lya\ flux. The second method is more in-depth and involves calculating the differential emission measure (DEM) for each of the transitions we measure and from there backing out the flux from transitions we do not measure.

\subsubsection{Method 1: Scaling from UV line fluxes}\label{subsubsec:method1}

\begin{figure*}
\includegraphics[width=\textwidth]{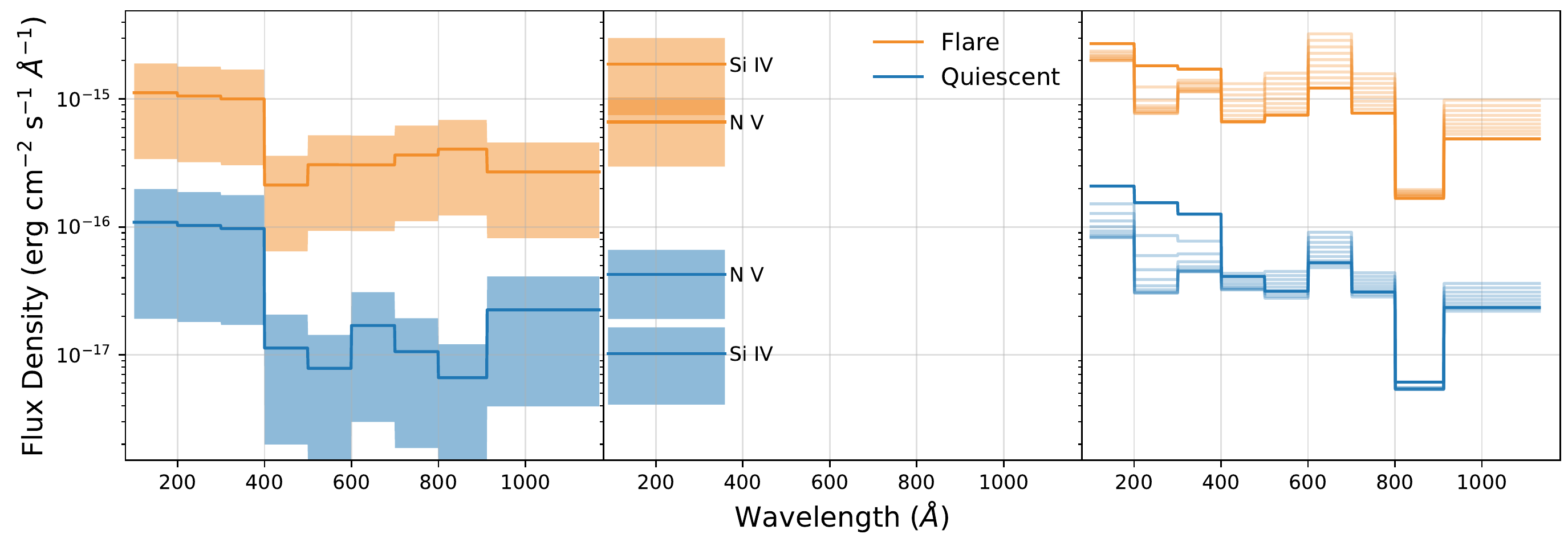}
\caption{The EUV estimates of \LHS\ flux for the quiescent (blue) and flare (orange) spectra. In the left panel, the EUV flux estimate is  scaled from the estimated \Lya\ flux \citep[Section~\ref{subsubsec:method1};][]{Linsky2014}. The lines show the best EUV estimates, and the shaded regions indicate 1$\sigma$ uncertainties. In the middle panel, the EUV flux from 90 ti 360 \AA\ is estimated using the measured Si \textsc{iv} and N \textsc{v} emission in the HST/COS data \citep[Section~\ref{subsubsec:method1};][]{France2018}. The lines and shaded regions are again the best estimates and 1$\sigma$ uncertainties. The right panel shows the EUV flux estimated from the DEM functions (Section~\ref{subsubsec:method2}). We intentionally used the same binning as the \Lya\ scaling (left panel) for comparison. Lighter blue and orange lines demonstrate the range of EUV estimates that arises if the assumed soft X-ray flux is 1 to $1\times 10^{-9}$ times the derived upper limit for \LHS\ from Swift-XRT (see Section~\ref{subsec:Xrayupperlimit}). }
\label{fig:EUVestimatesScaling}
\end{figure*}

We use UV--EUV scalings to estimate \LHS's EUV flux from its UV spectrum. The first scaling we attempt is based on measurements and models of the solar spectrum, EUVE, and FUSE data and relates broad EUV bands to \Lya\ flux \citep{Linsky2014}. The second scaling focuses on the Si \textsc{iv} and N \textsc{v} lines in the FUV, measured by HST/COS and STIS, and relates them to the 90--360\AA\ range measured for active M dwarfs by EUVE \citep{France2018}. The Si \textsc{iv} and N \textsc{v} lines are particularly suited to estimating the EUV because they form in $\sim$10$^5$ K plasma, similar to the coronal region where the bulk of the EUV radiation is emitted \citep{Tilipman2021}. The Si \textsc{iv} and N \textsc{v} lines do not suffer from ISM extinction as do other FUV lines that form at lower temperatures, like C \textsc{ii} \citep{France2018}. The results of these scalings for both the quiescent and flare \LHS\ spectra are shown in the left and middle panels of Figure~\ref{fig:EUVestimatesScaling}.

We find good agreement between the \citet{Linsky2014} and \citet{France2018} methods where they overlap for the quiescent and flare spectra, respectively. We use the RMS scatter about the Si \textsc{iv} and N \textsc{v} fits to calculate the uncertainties in the estimated EUV flux, similar to how \citet{France2018} calculated the error bars in Figure 6 of their paper. Overall, these scaling methods are disfavored because they rely on scaling from either the \Lya\ line, which we do not directly measure, or two FUV lines, which are measured at low signal-to-noise ratio. Furthermore, these scalings are derived from measurements of active M stars, whereas \LHS\ is considered inactive. However, it is encouraging that there is general agreement between the EUV flux estimates. 

\subsubsection{Method 2: DEM}\label{subsubsec:method2}

The second method involves calculating the DEM of the detected UV transitions and using their formation temperatures to construct a smoothly varying function that allows us to estimate the fluxes of lines that we do not measure. This process is described in detail in \citet{Duvvuri2021}. This method assumes that the plasma from which these high-energy transitions emit photons is optically thin and in a state of collisional ionization equilibrium (i.e., de-excitation due to collisions is negligible). As such, it is necessary to also have the emissivity contribution function for each transition, which we calculate with the ChiantiPy package,\footnote{\href{https://chiantipy.readthedocs.io/en/latest/}{ChiantiPy v0.9.5}} which utilizes the CHIANTI atomic database, v9.0 \citep{Dere1997,Dere2019}. 

To calculate the DEMs for the UV emission lines we measure, we start with the equation 

\begin{equation}\label{eqn:DEM}
    I_\mathrm{ul} = \int_{T}G_\mathrm{ul}(T)\Psi(T)dT
\end{equation}

\noindent where $I_\mathrm{ul}$ is the intensity for each transition ($ul$ stands for the transition from an ``upper to lower'' energy level), $G_\mathrm{ul}(T)$ is the emissivity contribution function, $\Psi(T)$ is the DEM, and $T$ is the temperature range over which a particular transition occurs. We assume a solar-like electron density of $10^8$ when calculating the emissivity contribution function. The emissivity contribution functions $G_\mathrm{ul}(T)$ exhibit strong peaks in temperature where most of the flux is concentrated, meaning that the integral in Equation~\ref{eqn:DEM} reflects a narrow temperature range for a given transition region line. We place an estimate of the DEM at the peak formation temperature of each transition line by integrating over $G_\mathrm{ul}(T)$ and dividing the intensity by this value \citep[see Equation 8 of ][]{Duvvuri2021}.

To fill out the DEM function, we require X-ray data to inform the highest peak formation temperatures. Over two epochs in 2019, Swift-XRT observed LHS 3844 for a total of 31.8 ks (PI: Corrales), but it was too faint to detect. Instead, we report a 95\% confidence upper limit on the count rate of $2.3\times10^{-4}$ counts s$^{-1}$, calculated using the Bayesian method detailed in \citet{Kraft1991}. Assuming a single-temperature plasma at 0.25 keV, this count rate yields an upper limit on the unabsorbed observed flux of $1.1\times10^{-14}$ erg cm$^{-2}$ s$^{-1}$ for the 0.2--2.4 keV band. This gives an upper limit X-ray luminosity in the same band of $2.9\times10^{26}$ erg s$^{-1}$ and a fractional luminosity $L_{\mathrm{X}}/L_{\mathrm{Bol}}$ of $3.1\times10^{-5}$ assuming $L_{\mathrm{Bol}}=9.34\times10^{30}$ erg s$^{-1}$.

In order to translate the upper limit unabsorbed flux into a DEM, we again start with Equation~\ref{eqn:DEM}. However, where $I_\mathrm{ul}$ previously applied to an individual transition that we measured, now $I_\mathrm{ul}$ represents the X-ray intensity across 0.2--2.4 keV (5.2--62.0 \AA). And instead of using $G_\mathrm{ul}(T)$ for a particular transition, we now sum the emissivity contribution functions for every ion transition in the band. We take the peak of this summation of $G_\mathrm{ul}(T)$ functions as the peak formation temperature for the X-ray band. We integrate the summed $G_\mathrm{ul}(T)$ function and divide the upper limit X-ray intensity by this value to get an estimate of the X-ray DEM, which serves as an upper limit given the nondetection by Swift-XRT. 

\begin{figure*}
\includegraphics[width=\textwidth]{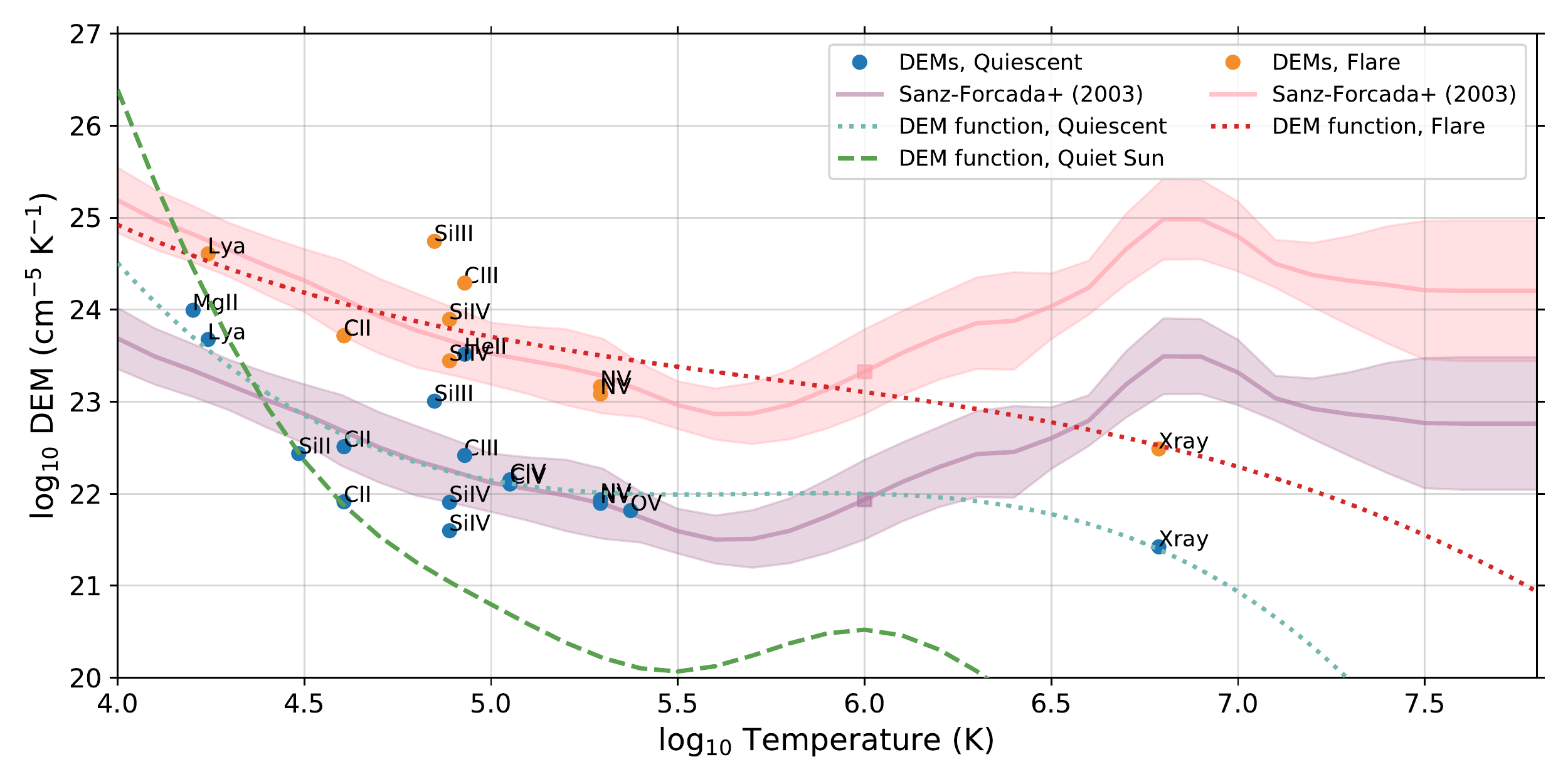}
\caption{The DEMs for observed transition lines in the HST/COS UV spectrum of \LHS. The X-ray data points are upper limits derived from Swift-XRT observations.  Blue points are DEMs from the quiescent spectrum, while orange points are DEMs from the flare spectrum. We compare the quiescent DEMs to those observed for active G stars by \citet{Sanz-Forcada2003}, with purple and pink lines and 1$\sigma$ error shading taken from an average across all stars in their sample, and then scaled to the quiescent and flare DEMS, respectively. The DEM function of the quiet Sun is plotted as a green dashed line \citep{Dere2019}. The teal and red dotted lines are third-order polynomial fits to the quiescent and flare DEM points, respectively. We include a point from the \citet{Sanz-Forcada2003} scaled DEM function at $\mathrm{log}(T)=6.0$ (purple and pink squares for the quiescent and flare DEMs, respectively) in the unconstrained area of parameter space between $\mathrm{log}(T)=5.5$ and $\mathrm{log}(T)=6.5$.}
\label{fig:EUVestimatesDEM}
\end{figure*}

We present our DEMs as a function of peak formation temperature in Figure~\ref{fig:EUVestimatesDEM}. We compare the DEMs from the quiescent line transitions to a scaled average of DEM functions across the stellar sample reported by \citet{Sanz-Forcada2003}, which was a strategy used by \citet{Garcia-Sage2017} for the $4.0<\mathrm{log}(T)<6.25$ temperature range of Proxima Centauri. However, the stars used in the \citet{Sanz-Forcada2003} study were mostly active, Sun-like stars, unlike \LHS, a relatively small and inactive star \citep[$R=0.19R_{\odot}$;][]{Vanderspek2019}. We therefore do not adopt the averaged \citet{Sanz-Forcada2003} DEM function for LHS 3844. Rather, we include a point from the \citet{Sanz-Forcada2003} scaled average DEM at $\mathrm{log}(T)=6.0$ where we have no constraining data, and then fit a third-order polynomial to the DEMs \citep{Louden2017}. For the flare spectrum, where we have fewer measured emission lines and no soft X-ray upper limit, we scale the \citet{Sanz-Forcada2003} DEM function estimate to the measured FUV DEMs and then use this scaling to increase the soft X-ray upper limit before fitting the polynomial. 

From the estimated DEM functions, we can back out the fluxes of lines that we do not directly measure, for example, in the range from 100 to 912 \AA. This simply requires going back to Equation~\ref{eqn:DEM} and using the DEM function and the emissivity contribution functions $G_\mathrm{ul}(T)$ from CHIANTI to calculate the intensity for each transition. We take the top 10 transitions of every ion in the CHIANTI database from 1 to 3215 \AA\ and compute the associated fluxes. We then insert these computed fluxes for high-energy wavelength ranges in which we do not have measurements of \LHS\ from either HST/COS (1131--3215 \AA) or Swift-XRT (5.2--62.0 \AA). The DEM estimates of the EUV flux are in agreement, to at least within an order of magnitude, with other EUV flux estimates we attempted (Figure~\ref{fig:EUVestimatesScaling}, right panel). We note that by using the DEM to estimate the EUV flux, we do not include an estimate on the Lyman continuum emission, which peaks at 911\AA. In the Sun, the Lyman continuum accounts for 15\% of the total EUV flux \citep{Woods2009,Duvvuri2021}.

We do not have data outside of the G130M grating for the flare spectrum. We instead use the flare DEM function to estimate the flux densities of the prominent lines that we observe in quiescence (C~\textsc{iv}, He~\textsc{ii}, Mg~\textsc{ii}). For these lines, we insert Gaussian profiles into the flare spectrum that integrate to the estimated values. We do not insert Gaussian profiles for lines that fall within the G130M grating but are not detected above the noise (Si~\textsc{ii}, O~\textsc{v}).

\subsection{Optical and infrared}

\begin{figure*}
\includegraphics[width=\textwidth]{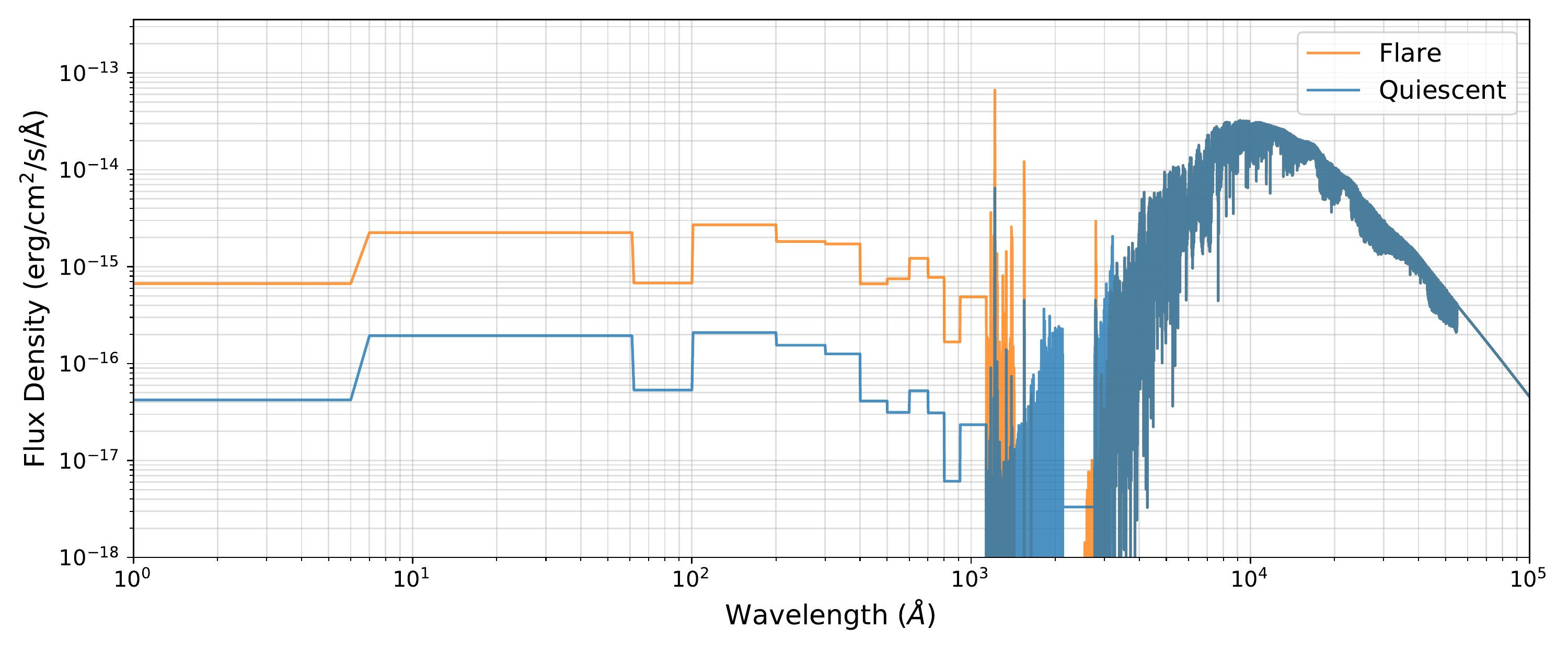}
\caption{Quiescent (blue) and flare (orange) spectra of \LHS\ from 1 to $1\times10^5$\AA. These panchromatic spectra combine UV transition region lines measured with HST/COS, an upper limit on the soft X-ray flux, an estimate of the EUV (100-912\AA) using a DEM, a PHOENIX model \citep{Husser2013}, and a blackbody curve. The flare spectrum is only directly measured in the FUV (see Section~\ref{sec:flare}), and the rest is estimated. There is a gap in the quiescent NUV observations between 2125 and 2750\AA, which we fill in with an average of the flux from 300\AA\ on either side of the gap, excluding the Mg \textsc{ii} emission lines.}
\label{fig:modifiedUV}
\end{figure*}

Redward of 3215 \AA\ we use a PHOENIX model that has been interpolated to the effective temperature and surface gravity of \LHS\ \citep{Husser2013}. We check that the interpolated model agrees with the available photometry. The PHOENIX model extends out to 5 $\mu$m. Past this, up to 10 $\mu$m, we append a blackbody curve at \LHS's effective temperature. We scale the combined PHOENIX and blackbody spectrum to the distance and stellar surface area of \LHS. We then compare the bolometric flux of the scaled PHOENIX and blackbody spectrum to $F_{\mathrm{bol}}=3.52\times10^{-10}$ erg cm$^{-2}$ s$^{-1}$ for \LHS\ reported by \citet{Kreidberg2019} from an SED fit. We further scale the spectrum such that it integrates to this $F_{\mathrm{bol}}$ value. (The scale factor is 1.01, so this is not changing the spectrum significantly.) We check that the scaled PHOENIX spectrum matches, to within the error bars, the COS spectrum at the reddest NUV wavelengths, where we detect a small amount of continuum flux from \LHS. We then append the optical and infrared portions of the spectrum to the end of the HST/COS data set. The panchromatic quiescent and flare spectra of \LHS\ are shown in Figure~\ref{fig:modifiedUV}.

\subsection{\LHS\ compared to MUSCLES stars}

The MUSCLES survey features few stars of similar spectral type to \LHS\ \citep{France2018}. The upcoming extension to MUSCLES for low-mass stars, Mega-MUSCLES (PI: Froning), will focus on this parameter space and so far includes Barnard's Star \citep{France2020} and the late-M dwarf TRAPPIST-1 \citep{Wilson2021}. We present some quantitative qualities of the \LHS\ spectrum in Table~\ref{tab:values}. We find that the quiescent \LHS\ spectrum fits within broad trends found for the MUSCLES stars. For instance, the FUV/NUV ratio for \LHS\ follows the trend toward higher FUV flux relative to NUV flux when going from warmer to to cooler stellar effective temperatures \citep{France2016}. The estimate of the quiescent \Lya\ line makes up $75\%\pm18$\% of the FUV flux, which is in agreement with MUSCLES stars. We note that our estimate of the XUV flux is higher than that of the MUSCLES stars. This is due to the differences between using the \citet{Linsky2014} scalings from the \Lya\ flux \citep{France2016} and a DEM function, similar to what we see for the EUV in Figure~\ref{fig:EUVestimatesScaling}.

\begin{deluxetable}{l|ccc}
\centering
\caption{Derived Values from the Spectrum of \LHS\label{tab:values}}
\tablewidth{0pt}
\tablehead{
\colhead{} & \colhead{Wavelength (\AA)} & \colhead{Value\tsup{a}} & \colhead{Value\tsup{b}}
}
\startdata
$\mathrm{log_{10}}(L_{\mathrm{Bol}})$ & 1 --- $1\times10^5$ & 30.97 & 30.97 \\
$f$(XUV) & 1 --- 912     & -3.65 & -2.49 \\
$f$(FUV) & 912 --- 1700  & -4.16 & -3.01 \\
$f$(NUV) & 1700 --- 3200 & -4.48 & -4.15 \\
\hline
FUV/NUV & --- & 2.11 & 13.81\tsup{c} \\
\Lya/FUV & --- & $0.74\pm0.61$ & $0.54\pm0.37$ \\
F$_{\mathrm{XUV,p}}$\tsup{d} & 1 -- 912 & $1.9\times10^4$ & $2.8\times10^5$\\
$\dot{M}$ (g s$^{-1}$) & --- & $7.7\times10^{10}$ & $1.3\times10^{12}$ \\
\enddata
\tablecomments{Following \citet{France2016}, $f$(band) =  log$_{10}$($L_{\mathrm{band}}/L_{\mathrm{Bol}}$); $F_{\mathrm{Bol}}$ is $3.52\times10^{-10}$ erg cm$^{-2}$ s$^{-1}$ \citep{Kreidberg2019} in all cases.\\
\tsup{a} Quiescent spectrum.\\
\tsup{b} Flare spectrum.\\
\tsup{c} We have no way of estimating the continuum NUV flux during the flare, so the integrated NUV flux is taken from the PHOENIX model and is therefore underestimated.\\
\tsup{d} The XUV flux (erg cm$^{-2}$ s$^{-1}$) at the planet \LHS b, assuming $a = 0.00622 \pm 0.00017$ au \citep{Vanderspek2019}
}
\end{deluxetable}

\section{Discussion} \label{sec:disc}

\subsection{The Soft X-ray Flux Upper Limit}\label{subsec:Xrayupperlimit}

\begin{figure}
\includegraphics[width=0.49\textwidth]{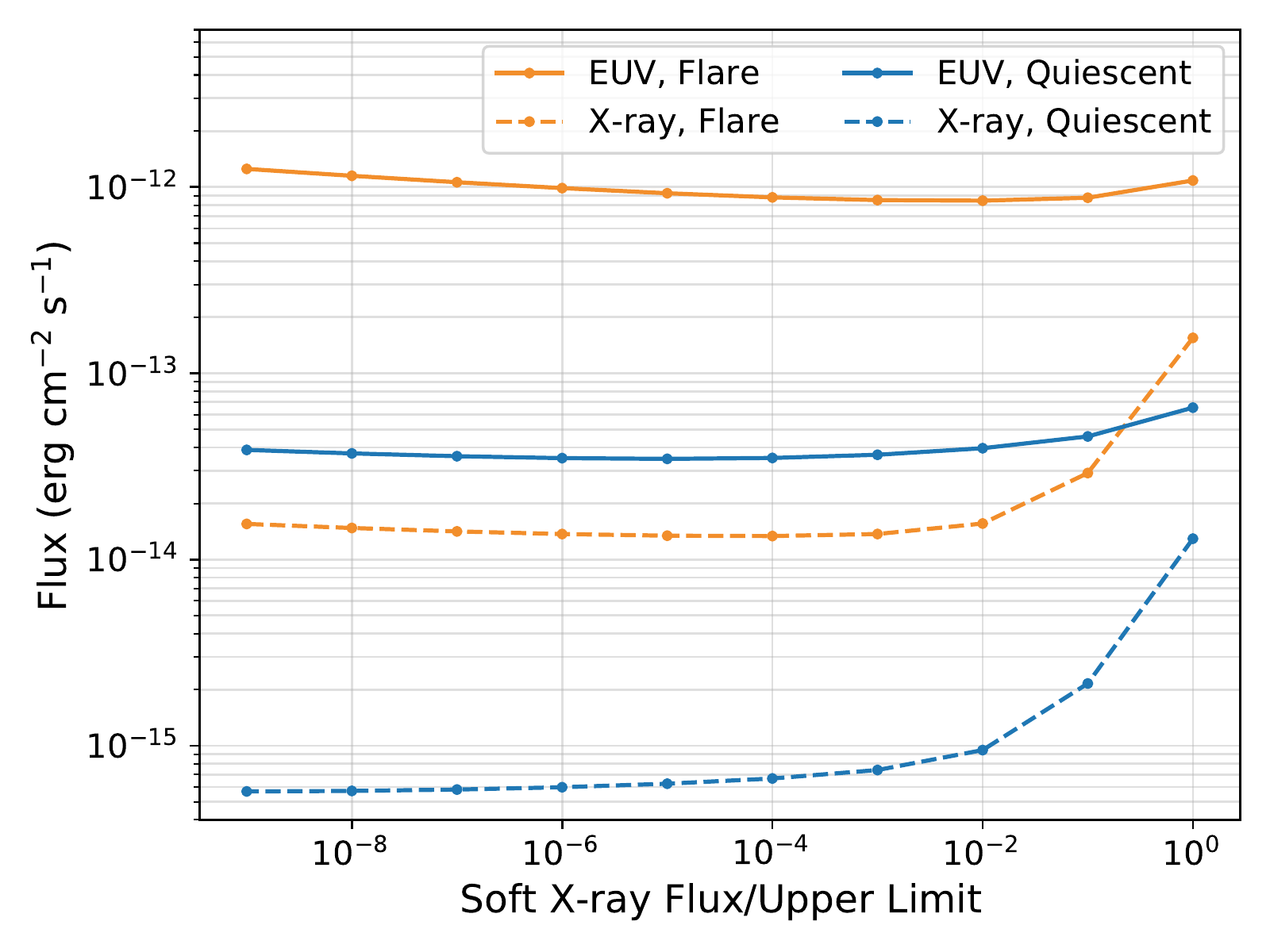}
\caption{Illustration of the response of the integrated X-ray (1--100 \AA) and EUV (100--912 \AA) flux to the assumed soft X-ray flux. From Swift-XRT, we derive an upper limit for the soft X-ray flux of $1.1\times10^{-14}$ erg cm$^{-2}$ s$^{-1}$. Here we take fractions of this upper limit and calculate how the integrated X-ray and EUV flux responds.}
\label{fig:X-raychange}
\end{figure}

Due to the nondetection of \LHS\ by the X-ray observatory Swift-XRT, we use an upper limit when constructing DEM functions for the quiescent and flare spectra presented in this work. Because the upper limit on \LHS's soft X-ray flux informs the highest peak formation temperatures of the quiescent and flare DEM functions (Figure~\ref{fig:EUVestimatesDEM}), we wish to understand the impact of taking the soft X-ray flux upper limit as the soft X-ray flux value. We systematically decrease the soft X-ray flux and recalculate the DEMs and resulting estimates for the integrated X-ray (1--100 \AA) and EUV (100--912 \AA) flux for \LHS. We present the results of this exercise in Figure~\ref{fig:X-raychange}, as well as illustrate the range of the resulting EUV flux estimates in  Figure~\ref{fig:EUVestimatesScaling} (right panel). As expected, changing the soft X-ray flux has a strong effect on the integrated X-ray flux, since the soft X-ray makes up a large portion of this spectral range. However, once the soft X-ray flux is decreased to 1\% of the upper limit, the integrated X-ray flux remains stable. In the EUV, the integrated flux changes by less than a factor of 2 in response to the decreased soft X-ray flux in both the quiescent and flare cases. The XUV (1--912 \AA) flux, which is dominated by the EUV, changes by less than a factor of 3. We are therefore comfortable taking the soft X-ray flux upper limit as the nominal value.

\subsection{Constraints on the Atmosphere of \LHS b}

The highly irradiated terrestrial planet in orbit around \LHS\ likely does not have an atmosphere. Atmospheres with surface pressures greater than 10 bars are disfavored \citep{Kreidberg2019}, and clear, low mean molecular weight atmospheres are disfavored down to 0.1 bar \citep{Diamond-Lowe2020b}. Tenuous high mean molecular weight atmospheres are not addressed empirically, but theoretical arguments based on the stellar wind and high-energy radiation that \LHS b receives strongly imply that this world is a bare rock \citep{Kreidberg2019}. 

\citet{Kreidberg2019} explored a parameter space for a final atmospheric pressure of O$_2$ on \LHS b after 5 Gyr of energy-limited atmospheric escape based on an initial water abundance and \LHS's XUV saturation fraction (Figure 4 of that paper). They explored an XUV saturation fraction from $10^{-4}$ to $10^{-2}$ and found a small area between 1 and $3\times10^{-4}$ XUV saturation fraction that allows for a high mean molecular weight atmosphere given the right initial water abundance. We can add more empirical information to this argument. From an upper limit on \LHS's soft X-ray flux and an estimation of the EUV, we find that in its current quiescent state, \LHS's XUV output flux relative to bolometric is $2.2\times10^{-4}$. 

With a rotation period of 128 days and no observed H$\alpha$ emission, \LHS\ is no longer in the saturation state exhibited by young stars \citep{Wright2018}. Models of activity evolution predict that M dwarfs spend an extended amount of time in the highly active pre-main-sequence phase before settling onto the main sequence \citep{Baraffe2002,Baraffe2015}, meaning that \LHS's XUV saturation fraction must have been higher than the fraction we see currently \citep[e.g.,][]{Luger2015}. The high-energy spectra we present in this work are representative of the unsaturated regime of \LHS\ and perhaps similar mid-M dwarfs as well.

\subsection{Mass-Loss Rate}

We can also use our XUV estimate to calculate an atmospheric mass-loss rate for \LHS b, supposing it had an extended low mean molecular weight atmosphere, as it may have soon after formation \citep{Owen&Wu2013,Lopez2013}. For planets that end up on the terrestrial side of the planet radius distribution gap \citep{Fulton2017,Fulton2018}, it is possible that they formed before the dissipation of the gas in the protoplanetary nebula. In this case, the terrestrial planet core would be able to accrete about 1\% of its mass in hydrogen and helium \citep{Owen2020}. Using the radius--mass relations of \citet{Chen2017}, we estimate that \LHS b currently has a mass of 2.2 $M_{\oplus}$ and so may have accreted 0.02 $M_{\oplus}$ of low mean molecular weight material soon after formation.

Basic mass-loss equations for energy-limited escape show that hydrodynamic escape from \LHS b would be rapid and efficient \citep[e.g., ][]{Erkaev2007,Lopez2013,Salz2016,Johnstone2019}. We calculate the evaporation efficiency log$_{10}(\eta_\mathrm{eva})$ to be -0.51 \citep{Salz2016}, placing \LHS b solidly in the regime of efficient atmospheric evaporation \citep[compare to Figure 2 of][]{Salz2016}. This is unsurprising, given that this is a terrestrial planet with a low gravitational potential. Following the equations and simulation results of \citet{Salz2016}, we calculate the mass-loss rate for the hydrodynamical regime using

\begin{equation}
    \dot M = \frac{3\beta^2\eta F_{\mathrm{XUV,p}}}{4KG\rho_{\mathrm{p}}}
\end{equation}

\noindent where we take $\eta=\eta_\mathrm{eva}$, $\beta$ relates the planet radius to the radius at which XUV flux can be absorbed, $F_{\mathrm{XUV,p}}$ is the XUV flux at the planet, $K$ is a correction factor because the mass only needs to reach the Hill radius to escape, and $\rho_\mathrm{p}$ is the planet density \citep{Erkaev2007,Salz2016}. At a distance of $0.00622 \pm 0.00017$ au from \LHS\ \citep{Vanderspek2019}, \LHS b receives $1.9\times10^4$ erg cm$^{-2}$ s$^{-1}$ in XUV flux when \LHS\ is in its current quiescent state. We find a hydrodynamic mass-loss rate of 7.7$\times 10^{10}$ g s$^{-1}$. At this rate, it would take only 54 Myr for \LHS b to lose its primordial low mean molecular weight atmosphere. We compare this mass-loss rate to that of the Neptune-sized world GJ 436b, which has a tail of hydrogen escaping at an estimated rate of 10$^8$-10$^9$ g s$^{-1}$ \citep{Ehrenreich2015}.

\subsection{Atmospheric Models}

While \LHS b is unlikely to retain an atmosphere, the panchromatic spectrum we present here for \LHS\ can be an input to atmospheric models of planets in inactive mid-M dwarf systems. For instance, we consider the case of a hypothetical Earth-like planet in the habitable zone of the \LHS\ system. We use the quiescent and flare \LHS\ spectra as inputs to the \texttt{Atmos}\footnote{ \href{https://github.com/VirtualPlanetaryLaboratory/atmos}{github.com/VirtualPlanetaryLaboratory/atmos}} coupled photochemical and climate model \citep{Arney2017} with updated opacities and molecular cross sections \citep{Grimm2015,Grimm2018,Gordon2017}. The nondetection of the UV continuum in the HST/COS data leads to negative flux density values in this part of the spectrum, which can create instabilities in the \texttt{Atmos} code. To avoid this, we create versions of the panchromatic spectra that are binned and resampled to 1 \AA\ such that negative flux density values are eliminated but the overall flux is conserved.

\begin{figure*}
\includegraphics[width=\textwidth]{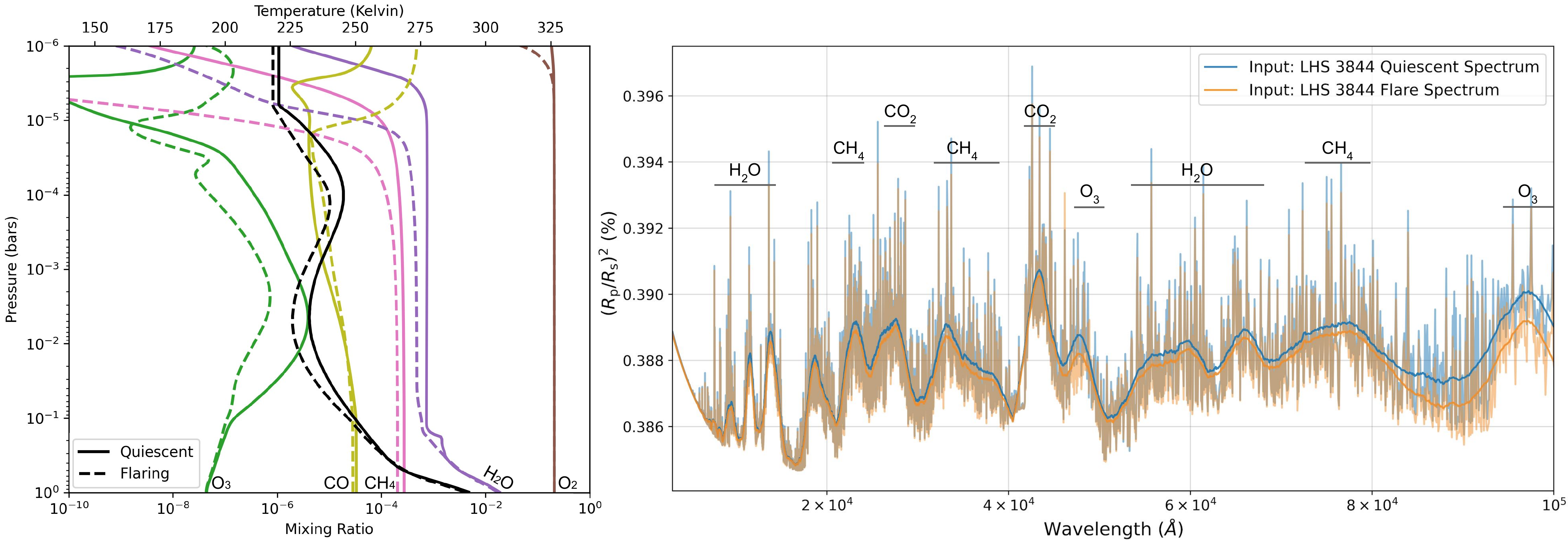}
\caption{Photochemical models for an Earth-like planet in the habitable zone of the \LHS\ system, with \LHS's quiescent and flare spectra as inputs. Left: T--P profiles for the quiescent and flare cases (black solid and dashed lines, respectively) along with mixing ratios for prominent molecules in Earth's atmosphere. Right: model transmission spectra derived from the photochemical models. The dominant species contributing to various spectral features are labeled.}
\label{fig:models}
\end{figure*}

We determine the steady-state temperature--pressure (T--P) profiles and mixing ratios for molecular species under both the quiescent and flare states of \LHS. The quiescent and flare spectra result in negligible differences in the T--P profiles, but yield significant differences in mixing ratios (Figure~\ref{fig:models}, left panel). During the flare, additional ozone (O$_3$) is produced in the upper atmosphere, while oxygen (O$_2$), water (H$_2$O), and methane (CH$_4$) are dissociated. 

The T--P and abundance profiles are fed into a proprietary modified version of the open-source \texttt{Exo-Transmit} code in order to produce model transmission spectra \citep{Kempton2017}. With this modification, we are able to read in vertically defined abundance profiles, as opposed to the equilibrium chemistry tables defined on preset T--P grids provided with \texttt{Exo-Transmit}. Investigating  the timescale of the planet's atmospheric response to the observed flare is beyond the scope of this work. We note that studies of energetic flares on active M dwarfs like AD Leo imply that they have a lasting affect on the steady state of a planetary atmosphere that may be detectable in transmission spectra \citep{Venot2016}. It is likely that \LHS\ exhibited heightened energetic flaring earlier in its lifetime, but it is not clear how long the relatively small flare we observed in this work would impact a planet's atmosphere. Based on the T--P profiles we produce for the hypothetical planet we consider here, we do not detect large differences in the resulting transmission spectra (Figure~\ref{fig:models}, right panel).

Based on studies of high-energy flares on active M stars, it is likely that low-energy stellar flares like the one observed in this work would leave atmospheric O$_3$ intact if it were to exist in a planetary atmosphere \citep{Segura2010,Tilley2019}. We are also not likely to see the effects of small stellar flares in the photochemical imprints they leave on planetary atmospheres, though it has been suggested that early stellar activity is a source of abiogenesis \citep{Ranjan2017,Rimmer2018}, which in turn could alter atmospheric compositions on geological times, as with the oxygenation of Earth by cyanobacteria. Atmospheric detections on terrestrial exoplanets are therefore best understood in the context of high-energy stellar radiation.

\section{Conclusion}\label{sec:conclusion}

We present the observed UV spectrum of \LHS\ from 1131 to 3215\AA\ using the G130M, G160M, and G230L gratings of the COS instrument  on board HST. During an exposure with the G130M grating, we detected a low-energy FUV flare. We separate the flare spectra from the quiescent spectra and present both in this work, providing two different pictures of \LHS. 

Because COS is a slitless spectrograph, we cannot directly observe the prominent \Lya\ line due to absorption by neutral hydrogen in the ISM and local geocoronal contamination. We instead use UV--UV scalings from the MUSCLES survey to estimate the \Lya\ flux \citep{France2016,Youngblood2017}. We estimate the EUV (100--912 \AA) flux using both the \citet{Linsky2014} and \citet{France2018} scalings, as well as a DEM function estimated from detected UV emission lines. We place an upper limit on the soft X-ray flux from Swift-XRT and include this in our estimates of the unmeasured high-energy portions of the \LHS\ spectrum. We find the DEM function estimates to be a more self-consistent method of estimating the unmeasured XUV (1--912\AA) flux of \LHS, an inactive mid-M star. However, we do find agreement to within an order of magnitude between the DEM and scaling methods where they overlap in wavelength space. We find that our resulting band-integrated measurements and estimates of the FUV, NUV, and XUV are in agreement with those found for other M stars \citep{France2013,France2016}.

We estimate that the highly irradiated exoplanet \LHS b currently receives $1.9\times10^{4}$ erg cm$^{-2}$ s$^{-1}$ in XUV flux and would lose a primordial hydrogen/helium atmosphere in 54 Myr. Keep in mind that this is the calculated hydrodynamic mass-loss rate derived from the current XUV flux of \LHS. This mid-M dwarf spent hundreds of millions of years in the the pre-main-sequence phase \citep{Reid2005}, and a further several gigayears in the X-ray saturation state \citep{Ribas2005,West2008,Stelzer2013,Wright2016,Wright2018}, during which its X-ray output was higher than it is today. While \LHS b is likely a bare rock \citep{Kreidberg2019,Diamond-Lowe2020b}, we present atmospheric models of a hypothetical Earth-like planet in the habitable zone of the \LHS\ system. Models of the molecular abundances of such a planet exhibit different responses to the \LHS\ quiescent and flare spectra. Though the atmosphere of this hypothetical planet would spend most of its time in the conditions presented for the quiescent star, the high-energy radiation of \LHS\ earlier in its lifetime would have looked more like the state we observe it in during the flare.

The forthcoming Mega-MUSCLES survey will place \LHS\ in the context of other planet-hosting mid-M stars. Assessing terrestrial exoplanet atmospheres requires knowledge of the high-energy environments provided by their host stars. It may not be possible to measure the UV spectrum of every planet-hosting M dwarf, and our ability to do so only lasts as long as HST can gather data. There is no planned successor on the scale of HST in the UV, though small satellites like the Colorado Ultraviolet Transit Experiment \citep{Fleming2018} and the Star-Planet Activity Research CubeSat \citep{Scowen2018} aim to measure the high-energy radiation of bright stars. There are also efforts to broadly capture the UV behavior of M dwarfs from optical indicators such as Ca \textsc{ii} and H$\alpha$ \citep{Melbourne2020}. Refining the high-energy behavior of mid-M dwarfs is crucial for future efforts to understand the atmospheres of the terrestrial worlds around them.

\acknowledgments
This paper includes data gathered with the Cosmic Origins Spectrograph on board the Hubble Space Telescope. We thank HST Program Coordinator Patricia Royle for assistance in scheduling these observations. We thank Amber Medina for valuable conversations about stellar flares. We greatly appreciate the thoughtful comments provided by the anonymous referee. Support for Program 15704 was provided by NASA through a grant from the Space Telescope Science Institute, which is operated by the Association of Universities for Research in Astronomy, Inc., under NASA contract NAS5-26555. We also make use of the CHIANTI database and open-source python code. CHIANTI is a collaborative project involving George Mason University, University of Michigan (USA), University of Cambridge (UK), and NASA Goddard Space Flight Center (USA). E.M.-R.K. and D.J.T.  recognize support from STScI program No.\ 16135 and NSF grant No.\ 2009095. This publication was made possible through the support of a grant from the John Templeton Foundation. The opinions expressed here are those of the authors and do not necessarily reflect the views of the John Templeton Foundation.\\

\facilities{HST(COS), Swift(XRT)}.

\software{\texttt{astropy} \citep{AstropyCollaboration2013,AstropyCollaboration2018}, \texttt{Atmos} (\href{https://github.com/VirtualPlanetaryLaboratory/atmos}{github.com/VirtualPlanetaryLaboratory/atmos}), \texttt{ChiantiPy} \citep{Dere1997,Dere2019}, \texttt{emcee} \citep{Foreman-Mackey2013}, \texttt{lmfit} \citep{Newville2016}}

Data products are available as HLSPs at MAST via \dataset[10.17909/t9-fqky-7k61]{\doi{10.17909/t9-fqky-7k61}}.

\bibliography{MasterBibliography.bib}

\end{document}